\documentclass[%
tikz,
showpacs,
preprintnumbers,
 amsmath,amssymb,
aps,
longbibliography
]{revtex4-2}

\usepackage{dcolumn}
\usepackage{bm}
\usepackage[dvips]{dropping}
\usepackage{hyperref}
\hypersetup{
    colorlinks = true,
    linkcolor = {blue},
    citecolor  = {blue},
    filecolor = {blue},
    menucolor= {blue},
    runcolor = {blue},
    urlcolor = {blue},
    linkcolor = {blue}
}

\usepackage{svg}
\usepackage{psfrag}
\usepackage{latexsym}
\usepackage{amsmath}
\usepackage{multirow}
\usepackage{amssymb}
\usepackage{epsfig}
\usepackage{graphicx}
\usepackage{array}
\usepackage{xfrac}
\usepackage{bm}
\usepackage{bigints}
\usepackage{color}
\usepackage{tikz}
\usepackage{pifont}
\usepackage{mathtools}
\usetikzlibrary{decorations}
\usetikzlibrary{calc,decorations.pathreplacing}
\usetikzlibrary{plotmarks}
\usepackage{flushend}
\usepackage{upgreek}
\usepackage{scrextend}
\usepackage{lipsum}
\usepackage{pgfplots}
\usepackage{rotating}
\usepackage{cancel}
\usepackage{booktabs}
\usepackage{tabularx}
\usepackage{tikz}
\usepackage{varwidth}
\usepackage{float}
\usetikzlibrary{positioning,arrows.meta,quotes}
\pgfplotsset{compat=1.18} 

\newcommand{\revision}[1]{{\color{black}#1}}

\usepackage[paperwidth=7in,paperheight=10in, textwidth = 139mm, textheight = 8.2in]{geometry}
\usepackage{enumerate}%
\usepackage{siunitx}

\begin{document}

\definecolor{mycolor1}{rgb}{0.3020,    0.6863,    0.2902}%
\definecolor{mycolor2}{rgb}{0.2157,    0.4941,    0.7216}%
\definecolor{mycolor3}{rgb}{0.8941,    0.1020,    0.1098}%
\definecolor{mycolor4}{rgb}{0.7725,    0.0667,    0.3529}%
\definecolor{mycolor5}{rgb}{1,    0.6501,    0}%
\definecolor{mycolor6}{rgb}{0	0.0667	0.3608}%

\definecolor{mycolorGrey}{rgb}{0.5, 0.5, 0.5}%

\newcommand{\lineStyleRef}{\raisebox{2pt}{\tikz{\draw[-,black,solid,line width = 1.5pt](0,0) -- (7mm,0)}}}

\newcommand{\lineStyleA}{\raisebox{2pt}{\tikz{\draw[-,mycolor1,solid,line width = 1.5pt](0,0) -- (7mm,0)}}}

\newcommand{\lineStyleB}{\raisebox{2pt}{\tikz{\draw[-,mycolor2,solid,line width = 1.5pt](0,0) -- (7mm,0)}}}

\newcommand{\lineStyleC}{\raisebox{2pt}{\tikz{\draw[-,mycolor3,solid,line width = 1.5pt](0,0) -- (7mm,0)}}}

\newcommand{\lineStyleRefshort}{\raisebox{2pt}{\protect \tikz{\protect \draw[-,black,solid,line width = 1.5pt](0,0) -- (4mm,0)}}}

\newcommand{\lineStyleAshort}{\raisebox{2pt}{\protect \tikz{\protect \draw[-,mycolor1,solid,line width = 1.5pt](0,0) -- (4mm,0)}}}

\newcommand{\lineStyleBshort}{\raisebox{2pt}{\protect \tikz{\protect \draw[-,mycolor2,solid,line width = 1.5pt](0,0) -- (4mm,0)}}}

\newcommand{\lineStyleCshort}{\raisebox{2pt}{\protect \tikz{\protect \draw[-,mycolor3,solid,line width = 1.5pt](0,0) -- (4mm,0)}}}

\newcommand{\lineStyleSTW}{\raisebox{2pt}{\protect \tikz{\protect \draw[-,mycolor4,solid,line width = 1.5pt](0,0) -- (4mm,0)}}}

\newcommand{\lineStyleOW}{\raisebox{2pt}{\protect \tikz{\protect \draw[-,mycolor5,solid,line width = 1.5pt](0,0) -- (4mm,0)}}}

\newcommand{\lineStyleSW}{\raisebox{2pt}{\protect \tikz{\protect \draw[-,mycolor6,solid,line width = 1.5pt](0,0) -- (4mm,0)}}}

\newcommand{\lineStyleViotti}{\raisebox{0pt}
{\protect\tikz{\protect\draw[-,mycolorGrey,solid,line width = 1.5pt](0,0) -- (7mm,0); 
\protect\draw[color=mycolorGrey, fill=mycolorGrey] plot[mark=*, mark size=2pt] (3.5mm,0)
}}}

\newcommand{\lineStyleGQ}{\raisebox{0pt}
{\protect\tikz{\protect\draw[-,mycolorGrey,dotted,line width = 1.5pt](0,0) -- (7mm,0); 
\protect\draw[color=mycolorGrey, fill=mycolorGrey] plot[mark=triangle*, mark size=3pt] (3.5mm,0)
}}}

\preprint{APS - Physical Review Fluids}

\title{Response of a turbulent boundary layer to steady, square-wave-type transverse wall-forcing}

\author{Max W. Knoop$^1$}
\email{m.w.knoop@tudelft.nl}

\author{Rahul Deshpande$^2$}
\author{Ferry F.J. Schrijer$^1$}
\author{Bas W. van Oudheusden$^1$}

\affiliation{$^1$Faculty of Aerospace Engineering, Delft University of Technology, 2629HS Delft,
The Netherlands\\
$^2$Department of Mechanical Engineering, The University of Melbourne, Parkville, VIC 3010, Australia}
\date{\today}

\begin{abstract}
This study investigates the spatial evolution of a zero pressure gradient turbulent boundary layer (TBL) imposed by a square-wave (SqW) of 
steady spanwise wall-forcing, which varies along the streamwise direction ($x$). 
The SqW wall-forcing is imposed experimentally via a series of streamwise periodic belts running in opposite spanwise directions, following the methodology of \citeauthor{knoop_experimental_2024} (\emph{Exp Fluids}, vol 65, 2024), with the streamwise extent increased to beyond $\sim 11$ times the boundary layer thickness (${\delta}_o$) in the present study.
This unique setup is leveraged to investigate the influence of viscous-scaled wavelength of SqW wall-forcing on the turbulent drag reduction (DR) efficacy for $\lambda^+_x = $ 471 (sub-optimal), 942 (near-optimal), and 1884 (post-optimal conditions), at fixed viscous-scaled wall-forcing amplitude, $A^+ = 12$, and friction Reynolds number, $Re_\tau = 960$.
The TBL's response to this wall-forcing is elucidated by drawing inspiration from established knowledge on traditionally studied sinusoidal forcing (SinW), based on analysis of the streamwise-phase variation of the Stokes strain rate (SSR).
The analysis reveals the SqW forcing to be characterized by a combination of two markedly different SSR regimes whose influence on the overlying turbulence is found to depend on the forcing waveform: sub-phase-I of local and strong impulses of SSR downstream of the half- ($\lambda_x$/2) and full-phase ($\lambda_x$) locations, associated with a reversal in spanwise forcing directions, leading to significant turbulence attenuation, and sub-phase-II of near-zero SSR over the remainder of forcing phase that enables turbulence recovery (when wall-forcing magnitudes and direction remain constant).

Upon the initial imposition of the SqW forcing, the Reynolds stresses are strongly attenuated over the short streamwise extent of $x/\delta_0<0.5$ for all wavelengths, whereas the skin-friction transient is more gradual. 
Thereafter, once the forcing is ultimately established, the sub-optimum and optimum wavelength regimes display no distinctive responses to the individual SSR sub-phases; rather, the drag-reduced TBL response is quasi-streamwise homogeneous.
In contrast, an SSR-related phenomenology establishes itself clearly for the post-optimal case, in which a local attenuation of near-wall turbulence characterizes sub-phase-I, while the turbulent energy recovers in sub-phase-II owing to the extended region of near-zero SSR.
\end{abstract}

\maketitle

\section{Introduction and motivation}
\label{intro}

\begin{figure}[t]
    \includegraphics{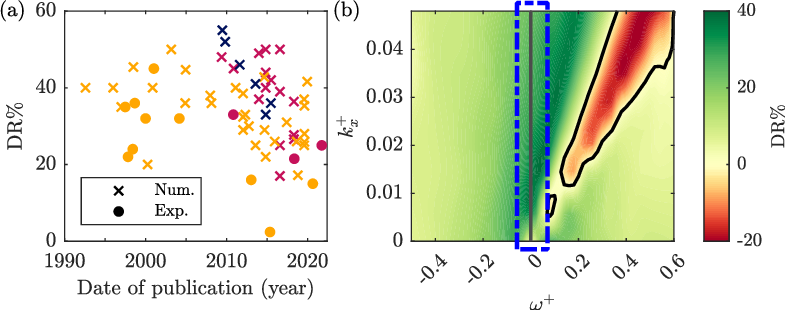}
    \caption{(a) Maximum DR noted by various investigations of spanwise wall-forcings published in the last three decades. 
    The forcing types are categorized according to streamwise traveling wave ($k_x$ $\neq$ 0, $\omega$ $\neq$ 0, \lineStyleSTW), time-oscillating ($k_x$ = 0, \lineStyleOW), and spatial forcing ($\omega$ = 0, \lineStyleSW). 
		The data was obtained from the review of \citet{ricco_review_2021}. 
    (b) DR noted for a turbulent channel flow showcased in the form of the `Quadrio map', for the aforementioned three different forcing types in the $k_x - \omega$ space ($\omega = 0$ has been highlighted by a dashed blue box). 
    The map is recreated from the statistics shared by \citet{gatti_reynolds-number_2016} for $A^+ = 12$ and $Re_\tau = 906$, which closely matches the present experimental conditions. Terminologies have been defined in \S\ref{intro}.}
    \label{fig:DR_Qmap}
\end{figure}

The ability to improve engineering efficiency, through manipulation of the turbulent skin-friction drag, has led to the proposal of several active and passive drag-reduction (DR) strategies in the literature.
One of the most promising active flow control strategies, in terms of DR magnitudes, involves the imposition of a streamwise wave of spatio-temporal spanwise wall oscillations \citep{ricco_review_2021}.
This strategy has been researched extensively over the past three decades across various possible implementations, as documented graphically in figure~\ref{fig:DR_Qmap}(a). 
This is primarily owing to its ability to yield significant DR with theoretical net power savings \citep{ricco_review_2021,marusic_energy-efficient_2021}.
Notwithstanding, the practical deployment of this strategy on engineering systems, such as on aircraft wings or within oil pipelines, faces numerous challenges owing to its complex system architecture. 
One avenue that holds potential is the passive recreation of the spanwise forcing, \emph{e.g.}, through geometrical surface deformations such as oblique wavy walls \cite{ghebali_large-scale_2017} or dimpled surfaces \cite{Nesselrooij_drag_2016}. 
Future design and/or optimization of such passive systems, however, can be significantly aided by understanding the fundamental flow physics over active wall-forcing scenarios, which forms the primary focus of this study.

The concept of streamwise traveling waves (STW), introduced by \citet{quadrio_streamwise-travelling_2009}, works by prescribing an oscillatory spanwise wall velocity following $W_w(x,t) = A\sin(k_x x - \omega t)$. 
Here $A$ is the forcing amplitude of spanwise wall velocity, $k_x$ and $\omega$ are the streamwise wavenumber and frequency, respectively, while $t$ denotes time.
We consider a Cartesian coordinate system where $x$, $y$, and $z$ denote the streamwise, wall-normal, and spanwise directions, with corresponding instantaneous velocity components represented by $U$, $V$, and $W$, respectively. 
The STW forcing has a rich DR response in the $k_x-\omega$ space depicted in figure~\ref{fig:DR_Qmap}(b), commonly referred to as the `Quadrio map'. 
The regime of downstream traveling waves ($\omega >0$) comprises a ridge of significant DR (in green), as high as 38.4\% for $A^+ = 12$ at $Re_\tau = 906$ \cite{gatti_reynolds-number_2016}. 
This is, however, flanked by a region of significant drag increase (in red), when the wavespeed $c = \omega/k_x$ increases to the order of the convection velocity of the near-wall turbulence (\emph{i.e.},  $\mathcal{U}_c^+ \approx 10$  \citep{kim1993propagation}). 
The superscript `+' here indicates viscous scaling based on the kinematic viscosity $\nu$ and the skin-friction velocity $U_\tau \equiv \sqrt{\tau_w/\rho}$, with $\tau_w$ being the wall-shear-stress and $\rho$ the fluid density. The friction Reynolds number is defined as $Re_\tau \equiv \delta U_\tau/\nu$, corresponding to the ratio of the boundary layer thickness ($\delta$) over the viscous length scale (${\nu}/{U_{\tau}}$).  

The STW reduces to a pure time-oscillating wall-forcing when $k_x = 0$, limiting the parameter space to the $x$-axis in figure~\ref{fig:DR_Qmap}(b). 
This forcing was first investigated using direct numerical simulations (DNS) by \citet{jung_suppression_1992}, almost 17 years prior to the introduction of the STW. 
They found a DR of 40\% on imposing forcing at $T^+ = 100$ ($T^+ = 2\pi/{\omega^+}$) and $A^+ = 12$ at $Re_{\tau} = 200$, which was later corroborated experimentally by \citet{laadhari_turbulence_1994}.
In general, it is evident from figure~\ref{fig:DR_Qmap}(a) that both STW ($k_{x}$ $\neq$ 0, $\omega$ $\neq$ 0) and time oscillating ($k_x$ $=$ 0) forcings have been investigated extensively in the literature, via both experimental and numerical approaches.
Interestingly, however, very little attention has been paid to the pure spatial forcing ($\omega$ = 0) scenario to date, having been considered by only six numerical studies (to the authors' knowledge), all of them limited to relatively low $Re_{\tau} \leq 1600$  \cite{viotti_streamwise_2009, yakeno_spatio-temporally_2009, skote2011turbulent, skote_comparison_2013, hurst2014effect, mishra_drag_2015}.
This forcing type corresponds to the vertical axis in figure~\ref{fig:DR_Qmap}(b), outlined by the blue dashed box, and forms the primary focus of this study.
Spatial forcings have been typically imposed in the past by prescribing a steady spanwise wall velocity, $W_w$, that varies harmonically with the streamwise distance:
\begin{equation}
    \label{eq:SW}
    W_w(x) = A\sin\left( {k_x}x\right) = A\sin\left( \frac{2\pi}{\lambda_x}x\right),
\end{equation} 
with ${\lambda}_x$ representing the streamwise actuation wavelength. 
Previous works of \citet{viotti_streamwise_2009} and \citet{yakeno_spatio-temporally_2009} have highlighted several similarities between the spatial and time-oscillating forcings.
For instance, the optimum DR for the former is found at $\lambda_x^+ \approx 1000$, which corresponds to $T^+$ $\approx$ 100 under a convective transformation ($\lambda^+_x = T^+ \mathcal{U}_c^+$) based on the nominal near-wall convection velocity mentioned previously ($\mathcal{U}_c^+ \approx 10$; \citep{kim1993propagation}).
However, the spatial forcing has been found to have a slightly higher DR potential than the time-oscillating forcing, with \citet{viotti_streamwise_2009} reporting DR $\sim$ 45\% for $A^+$ = 12 at $Re_{\tau}=200$. 
This can be understood based on figure~\ref{fig:DR_Qmap}(b), where the time-oscillating forcing ($k_x$ = 0) solely experiences a `footprint' of the high DR ridge. 
In contrast, the regime associated with spatial forcing  ($\omega$ = 0) `cuts' through the high DR ridge, offering a conservative alternative to the STW forcing by avoiding the drag-increasing regime altogether. 

\begin{figure}[t]
    \centering
    \includegraphics{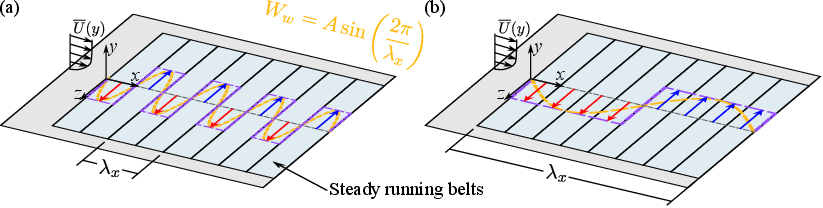}
    \caption{Schematic of the spatial wall-forcing imposed in the form of a steady sinusoidal wave (SinW; in orange) in past simulation studies, compared against steady square-wave (SqW; in purple) forcings achieved in experiments through discretized wall elements (marked in black). 
    Spatial forcing in (b) corresponds to a wavelength ${\lambda}_x$ that is four times that in (a). 
	Red and blue arrows, respectively, indicate positive and negative spanwise wall velocities ($W_w$).}
    \label{fig:SW_Forcing}
\end{figure}

A further benefit of a purely spatial forcing, compared to the STW forcing, also becomes apparent when considering the practical realizability of the associated wall-forcing setups.
Past experimental studies \citep{auteri_experimental_2010,bird_experimental_2018,chandran_turbulent_2023} investigating STW forcing considered discrete wall elements oscillating periodically along the spanwise/azimuthal direction at a particular $x$-location. 
This requires a much more complex system architecture than for a spatial forcing setup where the local spanwise wall velocity is constant (\emph{i.e.}, non-oscillatory).
This makes the latter more practically feasible when it comes to recreation through active (or passive) devices.
In the case of flat-plate turbulent boundary layers (TBL), for instance, spatial forcing can be imposed `actively' by using a series of steady spanwise running wall segments, similar to that realized in the study of \citet{kiesow_near-wall_2003}.
This concept is graphically depicted in figure~\ref{fig:SW_Forcing} where the idealized sinusoidal wave (SinW; equation~\ref{eq:SW}) forcing is shown, alongside its simplified and discretized version through a square-wave (SqW) forcing. 
The choice of the latter was motivated from the standpoint of practical realizability of such an experimental setup, as also adopted by past STW experimental studies \citep{auteri_experimental_2010,bird_experimental_2018,chandran_turbulent_2023}. 
Obviously, the SqW forcing does not affect the overlying turbulent flow in the exact same way as a SinW forcing, since it comes with a number of implications, which have been pointed out by past investigations \citep{auteri_experimental_2010, gallorini2024spatial}, namely: 
(i) the cumulative forcing intensity, defined by the average of the wall-velocity ($W_w$) magnitude, being higher for a SqW than a SinW for the same amplitude, 
(ii) an impulsive change in phase from positive to negative spanwise wall velocities in the case of a SqW.

A preliminary version of a wall-forcing setup, similar to figure~\ref{fig:SW_Forcing}, using four subsequent spanwise running belts, has been realized and validated previously by the present authors \citep{knoop_experimental_2024}.
The concept of this setup offers the unique capability to change the forcing wavelength ($\lambda_x$) independent of the forcing amplitude ($A$). 
This is made possible by the ability to vary $\lambda_x$ simply by changing the number of wall elements (\emph{i.e.}, belts) $S$, which make up one complete periodic waveform, while $A$ is manipulated by controlling the speed of the belts.
Figure~\ref{fig:SW_Forcing} schematically depicts the wavelength variation from (a) $S=2$ to (b) $S=8$, thereby permitting the forcing ${\lambda}_x$ for the latter to be four times that of the former.
This setup thus enables independent investigation of the influence of viscous-scaled forcing magnitudes ($A^+$) and forcing parameter (wavelength; $\lambda^+_x$) on the overlying TBL, which is otherwise challenging to achieve in the case of time-oscillating forcing setups where these parameters, $A^+$ and $\omega^+$ (or $T^+$), are commonly coupled.
Such investigations could be of particular interest in light of the recent work by Marusic and co-workers \citep{marusic_energy-efficient_2021, chandran_turbulent_2023,deshpande2024near}, which reported significant DR with theoretical net power savings for a high $Re_{\tau}$ TBL (of order $\mathcal{O}(\num{1e4}$)), based on STW forcings.
Their work adopted two different approaches of imposing STW forcings: the first involved imposing wall frequencies tuned to the small time-scale/large-frequency motions via the inner-scaled actuation (ISA) strategy (\emph{i.e.}, $\lambda_x^+ =\mathcal{O}(1000)$), while the second affected the large time-scales/small-frequency ($\lambda_x^+ \gg 1000$) motions via the outer-scaled actuation (OSA) strategy.

Forcing applied to a spatially developing TBL also presents the opportunity to study the initial transients at the onset of actuation, which allows us to further our understanding of the DR mechanism.
For instance, in the case of temporal wall-forcing, \citet{ricco_effects_2004} have shown experimentally that the DR exhibits a streamwise transient of $4-5\ \delta$, before reaching a more-or-less `fully developed' state in the TBL. 
In general, the available literature \citep{ricco_review_2021} suggests the relevant scaling parameter for the streamwise transient length to be $\delta$, rather than a forcing-related parameter (\emph{e.g.}, $T^+$, $\lambda_x^+$) or an inner scaling.
Such an initial spatial transient analysis, however, is not possible through conventional fully-developed channel flow simulations, which have been predominantly considered for investigating spatial wall-forcings in the past.
Further, while the majority of the past research has primarily focused on sinusoidal wave (SinW) forcing, the present study aims to explore, in particular, the streamwise transient response of a TBL to square-wave (SqW) wall-forcing, as motivated above by viewpoints from both fundamental (drag reducing mechanism) and practical (realizability) perspectives. 
To this end, particle image velocimetry is conducted across a large streamwise fetch of the zero-pressure gradient (ZPG) TBL, which is exposed to a steady spanwise wall-forcing along a SqW pattern in $x$.
Through these unique experiments, the paper aims to further our understanding of the DR mechanisms specific to SqW forcing, by documenting its phase-wise (spatial) turbulence modification, and contrasting it to that reported previously in the literature for SinW forcings \citep{Touber_near-wall_2012,agostini_turbulence_2015}.
Throughout this manuscript, we will be using `phase-wise' variation to correspond to the flow variation over the streamwise periodic domain: 0 $\le$ $x$ $\le$ $\lambda_x$. 

The present experiments will differentiate the spatial evolution of turbulence modification imposed via spatial wall-forcing in relation to wavelength variation, where we consider three regimes of sub-optimal, optimal, and post-optimal forcing.
As remarked previously, such an investigation is made possible via the independent variation of the actuation wavelength.
These insights would potentially be instrumental in developing passive forcing techniques in the future.

With regards to the organization of this manuscript, we first introduce the literature discussing the state-of-the-art understanding of achieving turbulence control and DR through SinW forcings, in \S\ref{sec:intro_mech}.
The details and methodology of our new experimental setup are then presented in \S\ref{sec:method}. 
The experimental results are documented in \S\ref{sec:results}, where we begin by analyzing the observed initial transient of the spatial modification of turbulence at the start of SqW forcing (\S\ref{sec:spatial_statistics}). 
Subsequently, in \S\ref{sec:analyticalSSR}, we provide a model to assess the rate of change of Stokes strain that was previously correlated with the phase-wise turbulence modification for a SinW. 
Hereafter, in \S\ref{sec:mechanisms}, we detail the implications of the established SqW-specific forcing on the modification of turbulence and skin friction at the initiation of forcing.
Our observations in relation to these forcing mechanisms are later substantiated in the domain where the wall-forcing is fully established (\S\ref{sec:downstream}). 
We conclude the results section by discussing the overall flow control efficacy of the SqW forcing via estimation of the nominal DR as a function of ${\lambda}^+_x$, in \S~\ref{sec:DR}. 
A conceptual overview of the flow phenomenology associated with SqW forcing is elaborated upon in \S\ref{sec:discussion}, followed by a discussion on its implications for the two waveform types and an outlook on the remaining related research questions of interest.

\subsection{The mechanism of drag reduction for sinusoidal transverse forcing}
\label{sec:intro_mech}

Numerous efforts have been made over the past decade to describe the flow physics underlying STW forcings \citep{Touber_near-wall_2012,Agostini_spanwise_2014,agostini_turbulence_2015,ricco_review_2021,rouhi2023turbulent}, but the exact mechanism behind their DR efficacy still remains largely unidentified. 
Here, we briefly review some literature that has hypothesized plausible DR mechanisms for spanwise forcings imposed via a SinW configuration. 
This will be used as a reference in \S\ref{sec:analyticalSSR} to compare and contrast with findings based on the SqW forcing configuration considered in the present study.
In the remainder of this manuscript, statistical quantities may be subjected to a triple decomposition of the instantaneous velocity components, corresponding to $U = \overline{U} + \widetilde{U} + u$ (and correspondingly for $V$ and $W$). 
The $\overline{\cdots}$ operator corresponds to the long-time averaged mean, whereas $\widetilde{U}$ denotes the phase-averaged quantities associated with the periodic forcing (temporal or spatial), and $u$ corresponds to the stochastic fluctuations. 
Similarly, the vorticity component associated with coordinate `$i$' is decomposed as $\Omega_i = \overline{\Omega_i} + \widetilde{\Omega_i} + \omega_i$. 

\begin{figure}[t]
    \centering
    \includegraphics{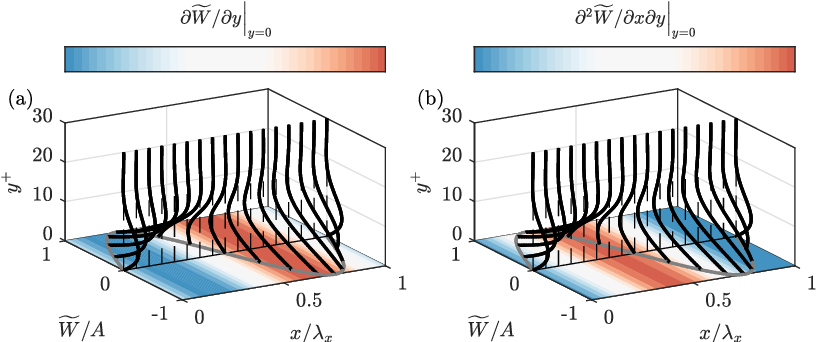}
        \caption{An example of the spanwise velocity profiles of the spatial Stokes layer for sinusoidal spanwise forcing. The grey lines denote the spanwise wall velocity ($\widetilde{W_w}$), and the black lines are representative of the wall-normal $\widetilde{W}$ profiles. The wall-based phase-wise variation of (a) the Stokes strain ($\left.\partial \widetilde{W}/ \partial y\right|_{y=0}$), and (b) the SSR ($\left.\partial^2 \widetilde{W}/ \partial y \partial x\right|_{y=0}$) have been visualized by the colored contours. The red and blue regions denote high SSR magnitude in (b), which corresponds with the negligible values of Stokes strain in (a). Vice-versa, near zero SSR is associated with a region of lingering (\emph{i.e.}, maximum) Stokes strain in (a).}.
    \label{fig:SSL_strain}
\end{figure}

Following the first numerical simulations of the time-oscillating forcing by \citet{jung_suppression_1992}, \citet{Akhavan_control_1993} and \citet{baron_turbulent_1996} proposed that the wall-normal gradient of the spanwise velocity ($\partial\widetilde{W}/\partial y$; \emph{i.e.}, the Stokes strain) is responsible for spatially decorrelating the low-speed streaks from the overlying quasi-streamwise vortices (QSVs). 
This decorrelation is a consequence of the reorientation of the near-wall velocity streaks into the direction of the mean strain (\emph{i.e.}, the vector $[\partial\overline{U}/\partial y\ 0\ \partial\widetilde{W}/\partial y]$), as was also noted by \citet{Touber_near-wall_2012}. 
It leads to the dampening of the self-sustaining cycle responsible for the majority of the near-wall turbulence kinetic energy production \citep{kim1993propagation,jimenez1999autonomous}, which was also noted experimentally by \citet{ricco_modification_2004} and \citet{kempaiah_3-dimensional_2020}. 
For reference, a phase-wise distribution of Stokes strain has been depicted by the colored contours in figure~\ref{fig:SSL_strain}(a), and compared against the corresponding spatial forcing; the wall-velocity $\widetilde{W_w}$ and its $\widetilde{W}$ profiles are shown in grey and black, respectively.
Imposition of the Stokes strain also results in a phase-averaged streamwise vorticity component ($\widetilde{\Omega_x} = \partial\widetilde{W}/\partial y$). 
\citet{Choi_mechanism_2001} hypothesized that $\widetilde{\Omega_x}$ gets periodically tilted by the imposed Stokes motion, giving rise to a (negative) net spanwise vorticity component ($\widetilde{\Omega_z}$) centered around $y^+ \simeq 15$.
By induction, this reduces the momentum at the wall (causing a drop in wall-shear stress) while increasing it in the upper part of the buffer layer (causing an overshoot of the mean streamwise velocity profile).
Several studies \cite{yakeno_spatio-temporally_2009, umair2023vorticity, miyake1997mechanism} have investigated the subsequent modification to the QSVs, which is central to the shear-stress producing mechanism \cite{orlandi1994generation}.
They found the QSVs to shift in the positive wall-normal direction while reducing their energetic content, thereby dampening the near-wall dynamics. 

Interestingly, recent studies from Leschziner and co-workers \citep{Touber_near-wall_2012,Agostini_spanwise_2014,agostini_turbulence_2015} have reported dampening of the near-wall streaks to be dependent on the rate of change in the Stokes strain.
\revision{The spatial forcing considered here can be quantified by the Stokes strain rate (SSR; $\partial^2 \widetilde{W}/ \partial y \partial x$), of which a phase-wise distribution at the wall has been depicted by the colored contours in figure~\ref{fig:SSL_strain}(b).}
For instance, \citet{agostini_turbulence_2015} highlighted the significance of the SSR in the manipulation of the near-wall vorticity dynamics, for post-optimal forcing at $T^+ = 200$. 
They hypothesized that the wall-normal vorticity is redistributed into spanwise vorticity in regions of strong SSR, through vortex stretching and tilting phenomena.
This redistribution of vorticity dampens the near-wall streaks, subsequently attenuating the QSVs and the wall-shear stress in accordance with the dynamics of the self-sustaining cycle \citep{jimenez1999autonomous}.
From a physical point of view, the regions of high SSR correspond to the phases where the spanwise shear strain reverses in sign.
Previous studies have found the near-wall streaks to dampen much more intensely during phases of high SSR magnitude, while the streaks tend to recover in the absence of Stokes strain variation (\emph{i.e.}, SSR $\approx$ 0). 
These interpretations are consistent with earlier investigations of constant straining conditions by \citet{Bradshaw_measuements_1985}, who associated the recovery in the near-wall turbulence to the re-establishment of an internal boundary layer, along the direction of the near-constant (\emph{i.e.}, `lingering') Stokes strain.
It is worth emphasizing here that the phase-wise variation of $C_f$ for post-optimal actuation does not `directly' respond to the imposed SSR; rather, \citet{agostini_turbulence_2015} report $C_f$ to increase in the region of high SSR at the wall, while it declines in the following lingering domain.
More recently, \citet{deshpande2024near} also investigated the lingering of the Stokes strain, in the case of STW forcings imposed on a high $Re_{\tau}$ TBL, by quantifying the instantaneous flow angle in the wall-parallel plane.
They found an increase in $T^+$ (which was coupled with reduced $A^+$) to result in the prolonged lingering of the flow angle, which correlated with the decrease in DR.
Analogous to the concept of SSR, \citet{ding2023acceleration} recently introduced an `acceleration' parameter ($a^+ = A^+/T^+$) for temporal SinW forcing, which was shown to scale the DR very well. 
The key takeaway from most recent studies on spatio-temporal wall-forcing, hence, is that the SSR and its phase-wise variation influence the turbulence modification and skin friction distinctly.
Later in \S\ref{sec:analyticalSSR} and \ref{sec:mechanisms}, we use this well-established understanding of the phase-wise variation of the SSR (for SinW forcing) to compare against novel observations for the SqW forcing, to explain the overlying turbulent flow modification.

\section{Experimental details}
\label{sec:method}

The experiments were performed in a subsonic open return wind tunnel at the Delft University of Technology, featuring a 4:1 contraction ratio. 
A flat plate turbulent boundary layer test section is employed downstream of the tunnel contraction, having a cross-section of $0.6\ \times\ 0.6 $ m$^2$ and a development length of $\sim2.5$\:m upstream of the actuation surface. 
The boundary layer is tripped to a turbulent state at the leading edge of the flat plate by a 0.12 m roughness strip of 40-grit sandpaper. 
A zero pressure gradient was ensured along the entire streamwise direction by adjusting the flexible top wall to the acceleration parameter \cite{Schultz_rough-wall_2007}, $K \equiv (\nu/U_e)^2(\mathrm{d}{U_e}/\mathrm{d}{x})$, such that a value of $K<\num{1.6e-7}$ is maintained under all operating conditions \cite{dacome_opposition_2023}, where $U_e$ is the velocity at the edge of the boundary layer.
A further detailed description of the boundary layer test section and its accompanying characteristics can be found in \citet{dacome_opposition_2023}.

\subsection{Spatial wall-forcing apparatus}
\label{sec:setup}
\begin{figure}[t]
    \centering
        \includegraphics{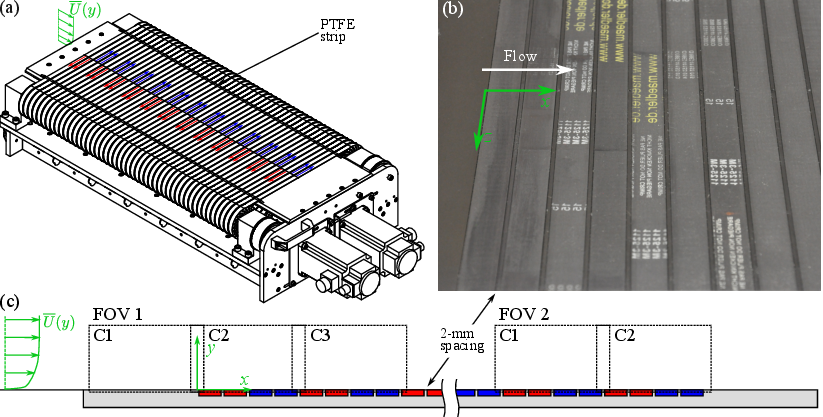}
    \caption{(a) Schematic of the experimental apparatus developed to impose spatial SqW wall-forcing. 
    (b) Photograph of the actuation surface/belts at the inflow. 
    (c) Schematic representation of the PIV setup in the streamwise-wall-normal plane. 
    The dotted boxes indicate the field-of-view (FOV) of individual cameras (C1$-$3), which are used to capture the flow in the upstream and downstream sections of the spatial wall-forcing setup. 
    Colored arrows and belts in (a,c) indicate positive (red) and negatively (blue) oriented $W_w$ vectors for a SqW waveform of $S=4$.}
    \label{fig:ExpSetup}
\end{figure}

This study deploys the concept of steady running belts in opposing spanwise direction \citep{kiesow_near-wall_2003}, discussed previously in \S\ref{intro}, for the experimental realization of the square-wave forcing setup.
While a preliminary version of this setup has been reported previously in \citet{knoop_experimental_2024}, here we present an extended version of the same concept that can affect the boundary layer for a much longer streamwise fetch.
The new spatial forcing setup (figure~\ref{fig:ExpSetup}a) comprises 48 belts extending across $\sim800$\:mm ($>$ 11$\delta_0$, with $\delta_0$ representing the TBL thickness for the non-actuated flow scenario (table \ref{tab:experimental_overview})) in the streamwise direction, making it $\sim$20 times longer than our previous setup.
This permits detailed investigations of the initial streamwise transients of the skin-friction drag and the overlying turbulent flow statistics along the wall-normal direction, followed by their steady-state/fully-developed characteristics. 
The actuating surface is comprised of neoprene belts with a streamwise extent of 15 mm and spanning $ \sim $ 294\: mm ($> 4\delta_0 $) in width on the test section floor. 
Subsequent belts are equally spaced along the streamwise direction with a spacing of 2 mm, which was the smallest possible distance realizable during setup fabrication.
Such a spacing was necessary to recess each belt into dedicated spanwise surface grooves, which ensured the actuation surface is flush with the wind tunnel flooring (schematically depicted in figure~\ref{fig:ExpSetup}c).
Strict tolerances were maintained on the actuation surface, with a maximum of 50\:{\textmu}m for the gap and step sizes.
Figure~\ref{fig:ExpSetup}(b) shows a photograph of the actuation surface at the inflow station, presenting an actual view of the belts embedded in the surface plate. 
Interested readers can also refer to a video recording of the actuated surface, which has been made available as supplemental material along with the manuscript \cite{SM1}.

The surface roughness of the neoprene belts is slightly higher compared to the upstream and downstream aluminum surface plate (\emph{i.e.}, tunnel floor).
However, it is within the tolerance limit for being considered as hydrodynamically smooth, with $k^+ \leq 5$ for the flow $Re_{\tau}$ under consideration \citep{jimenez2004turbulent}. 
The belts were constrained within the spanwise grooves (\emph{i.e.}, flush with the wall) by two PTFE strips that were bolted on each spanwise edge of the setup (figure~\ref{fig:ExpSetup}a). 
Despite this, some vibration and vertical displacement of the belts were observed to occur, which were characterized by a PSV-500 Poltec scanning vibrometer. 
For the actuation conditions in the current study, the worst scenario was found to be a 39\:{\textmu}m standard deviation of the wall-normal displacement, while the median displacement across the full actuation surface was 13\:{\textmu}m, both of which are within one viscous unit which is approximately 72 \textmu m. 
Regardless of the influence of the surface imperfection, a slight modulation of the near-wall turbulence was still observable in the final results (see \emph{e.g.}, figures~\ref{fig:FOV1_contours} and \ref{fig:FOV2_Cf_Rxx}).
These do not, however, affect our major findings/conclusions since we primarily contrast the flow physics for the actuated surface against those over the same surface in the non-actuated scenario.

Rotation to drive the belts is provided by two 6 Nm servo motors (LCMT-18L02), one for each spanwise direction, which can achieve a maximum spanwise velocity of 6 m/s. 
It is important to highlight the fact that the belts are driven in a steady fashion, following the prescription of a SqW spatial forcing sketched in figure~\ref{fig:SW_Forcing} previously. 
The motors are connected to two driving axes, including a non-slip pulley system that drives the belts directly. 
The pulleys can be connected or disconnected to the axis by employing a fixation mechanism, allowing interchangeability of the spanwise direction of the individual belts ($\pm$ $W_w$).
This unique feature permits the investigation of varying actuation/forcing wavelengths on the overlying TBL, as sketched previously in figure~\ref{fig:SW_Forcing}. 
We will denote the number of belt elements that make up a periodic waveform ($\lambda_x$) as `$S$'.
Considering the length of one belt `element' ($l_S$) as a combination of the belt's streamwise extent and periodic spacing/gap (\emph{i.e.}, $l_S = 17$\:mm), the actuation wavelength $\lambda_x = S\times l_S$. 
For convenience in the discussion of the results, we fix the origin of the coordinate system at 1\:mm upstream of the first belt.
On some occasions, we will discuss the integral effect of surface actuation by streamwise averaging of the considered flow property across the entire spatial phase (\emph{i.e.}, ${\lambda}_x$), which will be denoted by the $\langle ... \rangle_x$ operator.

\subsection{Particle image velocimetry}

\begin{table}[t]
    \centering
\begin{tabular}{ccccccccccccc}
        \hline \hline
\multicolumn{4}{c}{Inflow boundary layer characteristics}             & \multicolumn{5}{c}{Actuation characteristics}\\
\cmidrule(rl){1-4} \cmidrule(rl){5-9}
$U_\infty$ (m/s) & $U_{\tau0}$ (m/s) & $\delta_0$ (mm) & $Re_\tau$ & S   & $\lambda_x$ (mm) & $A$ (m/s) & $\lambda_x^+$ & $A^+$ &  line style\\
\hline
5          & 0.207              & 69.9            & 960      & $-$ & $-$                & $-$       & $-$           & $-$   &  \lineStyleRef \\
5          & 0.207             & 69.9            & 960      & 2   & 34                 & 2.46      & 471           & 12.0  &  \lineStyleA  \\
5          & 0.207              & 69.9            & 960      & 4   & 68                 & 2.46      & 942           & 12.0  &   \lineStyleB \\
5          & 0.207              & 69.9            & 960      & 8   & 136                & 2.46      & 1884          & 12.0  &  \lineStyleC \\  \hline \hline
\end{tabular}
    \caption{Overview of the experimental conditions and actuation parameters.}
    \label{tab:experimental_overview}
\end{table}

Multi-camera planar PIV measurements (2D-2C) were conducted in the streamwise-wall-normal plane ($x-y$) to quantify the corresponding velocity components in a wide field-of-view (FOV). 
An overlap of $\sim 15$\:mm between adjacent camera views was retained to allow for stitching of the individual vector fields.
Such wide FOV measurements were conducted both close to the start (FOV1) as well as towards the end (FOV2) of the actuation surface, as depicted schematically in figure~\ref{fig:ExpSetup}(c), independent of each other.
FOV1 captures nominally $\sim 70$\:mm of the inflow over the tunnel wall followed by that over belt\# $1-8$, using three sCMOS cameras (C1-C3) in a row, to image $\sim 210\ \times\ 49$ mm$^2$ in the streamwise and wall-normal directions, respectively.
This configuration enables investigation of the initial streamwise transients of the TBL flow for the wall-actuated cases.
On the other hand, FOV2 captures the flow field over belt\# $40-48$, employing two cameras (C1 \& C2) so as to image $\sim 140\ \times\ 49$ mm$^2$. 
As discussed later in \S\ref{sec:results}, the TBL drag and flow statistics are found to be nominally saturated at the belt locations in FOV2 and, as such, represent a fully developed regime of actuation.
This methodology provides the flow field statistics that permit quantification of the DR, as well as the turbulent stresses, for the various spatial forcing scenarios investigated in this study. 
While previous studies have investigated the energy and vorticity budgets to understand drag-reducing mechanisms \citep{Agostini_spanwise_2014,agostini_turbulence_2015,umair2023vorticity}, our current experimental, with its relatively large field of view (FOV) and the absence of higher spatial and temporal fidelity, prevents such detailed analysis. 
Henceforth, our primary focus is to characterize the streamwise evolution of the turbulent stresses to wall-forcing.

For data acquisition, digital LaVision sCMOS cameras ($2160\ \times\ 2560$ px$^2$, 16 bit, 6.5 $\mu$m) were used to acquire 2000 uncorrelated particle image pairs at a sampling frequency of 8 Hz ($1/{\Delta}t$).
The individual image pairs are, hence, separated by about 9 boundary layer turnovers times ($\delta_0/{{U_\infty}{\Delta}t}$). 
Nikkor AF-S 200 mm lenses were used at an f/5.6 aperture. 
A time separation between image pairs of 50\:{\textmu}s was selected to achieve particle displacements of $\sim$9 pixels in the freestream.
Resulting from the wall-motion, the maximum out-of-plane movement was $\sim0.12$\:mm, corresponding to about 1/8th of the laser sheet thickness. We consider this to be sufficiently within acceptable limits.
Illumination was provided by Quantel Evergreen 200 laser (Nd:Yag,  532\:nm, 200\:mJ) at 75\% power setting. 
Laser optics were employed to create a laser sheet of $\sim$1\:mm thickness. 
Seeding was generated using a SAFEX Fog 2010+ fog generator to inject 1 \textmu m water-glycol tracer particles into the air stream. 

The PIV images were processed using DaVis 10.2 software. 
It involved spatial filtering by subtracting a sliding Gaussian average, having a kernel width of 6 pixels, followed by a normalization with the local average over an 11-pixel kernel. 
Vector calculations were done using a cross-correlation algorithm, employing circular Gaussian interrogation windows of $16\ \times 16$ pixels at 75\% overlap. 
This resulted in the final spatial resolution of the PIV interrogation window to be of the order of 0.45 mm or 6.2 $\nu/U_{\tau0}$, where $U_{\tau0}$ is the friction velocity associated with the non-actuated flow.
The universal outlier detection algorithm \cite{Westerweel_outlier_2005} was utilized for vector validation, with a 2-times median filter to remove vectors with a residual above 2. 
Given the overlap, a vector pitch of 0.11 mm or 1.6 $\nu/U_{\tau0}$ was achieved. 

A total of four cases are discussed in the current study (refer table \ref{tab:experimental_overview}), namely a non-actuated case and three cases corresponding to varying streamwise actuation wavelengths: $\lambda_x^+ = {\lambda_x}{U_{\tau0}}/{\nu} = 471,\ 942$, and $1884$.
These wavelengths were achieved by changing the number of belt elements in the waveform to $S = 2,\ 4\ \textrm{and},\ 8$.
This wavelength range has been chosen strategically to cover the domain both greater and lower than the optimum of ${\lambda^+_x}$ $\approx$ 1000 \citep{viotti_streamwise_2009,yakeno_spatio-temporally_2009}.
The same inflow condition of tunnel freestream, $U_{\infty}$ $\approx$ 5 m/s, and a belt speed of 2.46 m/s, were maintained across all cases, resulting in a constant spatial forcing magnitude $A^+ = A/{U_{\tau0}} = 12$.

\subsection{Scaling of the mean velocity profiles and turbulent drag reduction}

Comparison between the mean velocity profiles for actuated and non-actuated cases can be made by the use of two different velocity scalings, either (i) the reference friction velocity $U_{\tau0}$ of the non-actuated case, or (ii) the actual friction velocity of the individual cases $U_\tau$. 
Throughout this study, we will follow the convention to denote normalization by the respective velocity scales via the superscripts `+' and `*'. 
It is now well-known \citep{gatti_reynolds-number_2016} that normalizing the mean velocity profile by reference $U_{\tau0}$ results in a collapse in the outer-layer, while using the actual $U_\tau$ yields a collapse in the viscous sublayer, along with an upward shift in the logarithmic layer (in case of DR). 

The large extent of the PIV FOV, combined with the measurement uncertainty in the near-wall region, results in a lack of reliable data in the viscous sublayer ($y^+ \leq 5$). 
Hence, accurate estimation of the skin friction directly from the gradient of the mean velocity profile ($\mathrm{d}\overline{U}/\mathrm{d}y|_{y=0}$) is not feasible. 
Instead, we can obtain $U_\tau$ indirectly for the actuated cases by fitting the mean velocity profiles to a modified composite profile, inspired from the original proposal of \citet{chauhan_criteria_2009}. 
To this end, we modify the composite formulation to accommodate for the additional $\Delta B$ shift that is expected to occur when normalization is made with the actual $U_\tau$ \citep{gatti_reynolds-number_2016}. 
Further details on the modified composite fitting procedures are outlined in appendix~\ref{app:profileFit}, where we discuss the efficacy of the method and implications of the modified fit.

Accordingly, the drag reduction (DR) can be expressed as the percentage difference of wall-shear stress with respect to the non-actuated case, which can also be expressed in terms of $U_\tau$ (assuming equal freestream velocity and density), according to:
\begin{equation}
    \label{eq:DR}
    \mathrm{DR\%} = \frac{\tau_{w0} - \tau_w}{\tau_{w0}}\times 100 = \left[1 - \left(\frac{U_\tau}{U_{\tau0}}\right)^2\right]\times 100. 
\end{equation}
Standard uncertainty propagation is applied to the $U_\tau$ components to obtain a 95\% confidence interval on the DR and the skin-friction coefficient, $C_f \equiv 2(U_\tau/U_\infty)^2$. 
In its canonical formulation, a narrow 95\% confidence interval $U_\tau$  of $\pm 0.7\%$ was found \cite{Rodriguez_robust_2015}. 
Similarly, we assessed the uncertainty of the modified fit from our actuated flow data in FOV2, which was found to be of similar order at $U_\tau \pm 0.75\% $ (refer appendix~\ref{app:profileFit} for further details). 
It is important to recognize that our evaluation of DR is not a direct measurement; as such, we do not treat the DR and $C_f$ as absolute values, but rather use them for purposes of qualitative assessment.
The qualitative trends are reproduced reliably through the aforementioned procedure, which is evidenced by their consistency with the existing literature (see, \emph{e.g.}, figures~\ref{fig:Umean_x}, \ref{fig:FOV1_linescomparison}, \ref{fig:FOV2_Cf_Rxx}).

\subsection{Inflow boundary layer characteristics}

\begin{figure}[t]
    \centering
    \includegraphics{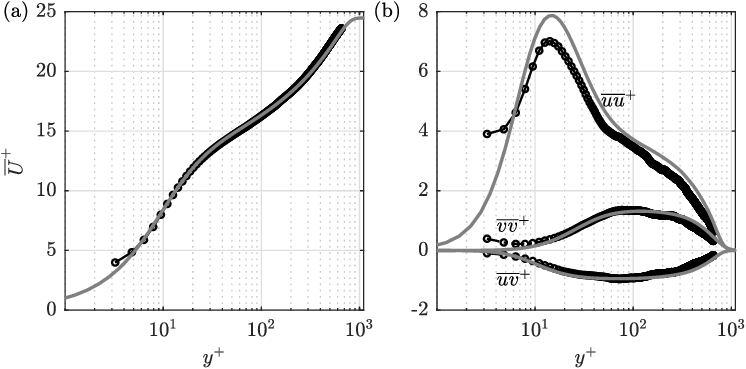}
    \caption{Wall-normal profiles of (a) mean streamwise velocity, $\overline{U}^+$ and the (b) Reynolds stresses $\overline{uu}^+$, $\overline{vv}^+$, and $\overline{uv}^+$ averaged across -11\:mm $\le$ $x$ $\le$ -1\:mm from the PIV FOV1, measured upstream of the actuation setup. 
    The grey solid lines are from the ZPG TBL DNS at $Re_\tau = 1060$ (with $\delta$ based on the composite fit) \citep{schlatter2010assessment}.}
    \label{fig:referenceBLprofiles}
\end{figure}

We begin by analyzing the turbulent boundary layer characteristics at the inflow upstream of the actuation surface to validate our PIV measurements and assess the canonical structure of the boundary layer. 
For this, we conduct ensemble and streamwise averaging of the velocity statistics to improve the statistical convergence of the data in the region $-11\ \mathrm{mm} \leq x \leq -1\ \mathrm{mm}$, corresponding to FOV1 (figure~\ref{fig:ExpSetup}c; $x$ = 0 is immediately upstream of the first belt). 
The friction velocity ($U_{\tau0}$) was found to be $U_{\tau0} \approx 0.207$ m/s via the composite profile method of \citet{chauhan_criteria_2009}, of which details are provided in \S \ref{app:profileFit}.
Since the present FOV1 does not extend across the entire TBL thickness in wall-normal direction, a separate PIV experiment was conducted (not discussed for brevity) to yield $\delta_0 = 69.9$\:mm and $Re_\tau = 960$.
Figures~\ref{fig:referenceBLprofiles} (a,b) depict profiles of the mean streamwise velocity and the Reynolds stresses from the PIV data, respectively (in black circles). 
These are compared against the profiles from the direct numerical simulation (DNS; in grey lines) of a ZPG TBL at a comparable $Re_\tau = 974$ \citep{schlatter2010assessment}. 
This confirms that the inflow TBL is canonical in nature, displaying only minor deviations from the DNS profiles, which are likely caused by the limitations of the PIV measurement technique. 
For instance, the disagreement of the near-wall $\overline{U}^+$ profile, below $y^+<4$, is attributed to biased velocity measurements owing to the PIV interrogation window partly overlapping the wall. 
The Reynolds stresses also exhibit a similar discrepancy in this region. 
The streamwise normal stress shows the largest deviation from the DNS reference, with an attenuation of the near-wall peak.
Moreover, a slight variation in the outer-layer is observed, attributed to the higher $Re_\tau$ of the DNS reference.
The disagreement in the near-wall region can be attributed to the attenuation of the small-scale energy due to finite-sized PIV interrogation windows and laser-sheet thickness, which is well-known from the literature \cite{lee2016validating}. 
The characteristic near-wall peak in the $\overline{uu}^+$ profile is nevertheless found at the appropriate location of $y^+ = 15$, demonstrating consistency with expected TBL statistics.
The limited dynamic range of PIV measurements for the $V$-component may introduce small-scale spurious fluctuations \cite{wilson2013uncertainty} resulting in a minor energy increase, plausibly explaining why the $\overline{uv}^+$ and $\overline{vv}^+$ components show no significant spatial attenuation despite expectations.
However, we are primarily concerned with larger-scale flow phenomena that are unaffected by these small scales and always contrast against a non-actuated reference; as such, these do not affect our conclusions.

\section{Results}
\label{sec:results}

\subsection{Streamwise evolution of turbulence statistics at the initiation of spatial forcing}
\label{sec:spatial_statistics}
\begin{figure}[t]
    \centering
    \includegraphics[width = \textwidth]{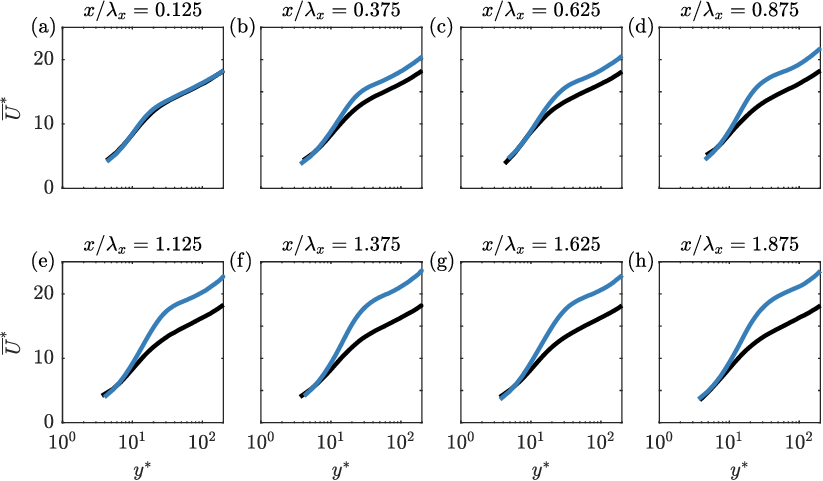}
    \caption{Downstream evolution of the mean velocity profile, following the onset of actuation.
			Comparing the non-actuated (in black) and actuated case ($\lambda^+_x = 942$ and $A^+ = 12$; in blue). 
			(a-h) correspond to the profiles over the respective belts\# $1-8$ from FOV1.}
    \label{fig:Umean_x}
\end{figure}

We investigate the initial streamwise transients of the actuated ZPG TBL over the upstream portion of the setup at $Re_\tau = 960$. 
For this, we consider PIV measurements from FOV1. This FOV covers a part of the inflow along with the first eight belts, which cover a streamwise extent of $\sim$2$\delta_0$, measured from the leading edge of the actuation surface. 
We compare first the non-actuated case to the near-optimum wall-forcing case, \emph{i.e.}, $\lambda_x^+ = 942$, having the maximum DR ~$\sim 38$\% (as will be quantified and discussed later in section~\ref{sec:DR}). 
The downstream development of $\overline{U}^*$ is first evaluated, with the skin friction obtained via the modified composite fit procedure (appendix~\ref{app:profileFit}).
Figures~\ref{fig:Umean_x}(a-h) present $\overline{U}^*$ corresponding to the center of each of the 8 belts in FOV1, which are obtained by spatially averaging $\overline{U}$ over a region of ${\Delta}x$ $\sim$ $\pm$0.007$\delta_0$ $\sim$ $\pm$6.9${\nu}/{U_{\tau_0}}$ 
Normalization of these profiles by the local $U_\tau$ enforces similarity in the inner region between the cases while causing a vertical shift (denoted by $\Delta B$) in the logarithmic layer, which is correlated to the degree of DR \citep{gatti_reynolds-number_2016}. 
For instance, the complete collapse of the non-actuated and actuated profiles over the first belt (figure~\ref{fig:Umean_x}a) suggests negligible DR at $x/{\lambda}_x$ = 0.125. 
However, a monotonic increase in ${\Delta}B$, \emph{i.e.}, increase in DR, can be observed with further downstream evolution. 
The drag changes particularly rapidly across the streamwise region: $0.375 \leq x/ {\lambda_x} \leq 1.125 $ (figures~\ref{fig:Umean_x}b-e), after which it tends to nominally saturate to a steady state level. 
These characteristics align with the literature \citep{ricco_modification_2004,Agostini_spanwise_2014}, wherein an initial streamwise transient is expected before the mean drag and turbulence flow statistics saturate/fully develop. 
The profiles also provide the reader with a qualitative sense of the accuracy of the modified composite fit, which underlines the upcoming discussion on the nominal streamwise variation of $C_f$.

\begin{figure}
    \centering
    \includegraphics{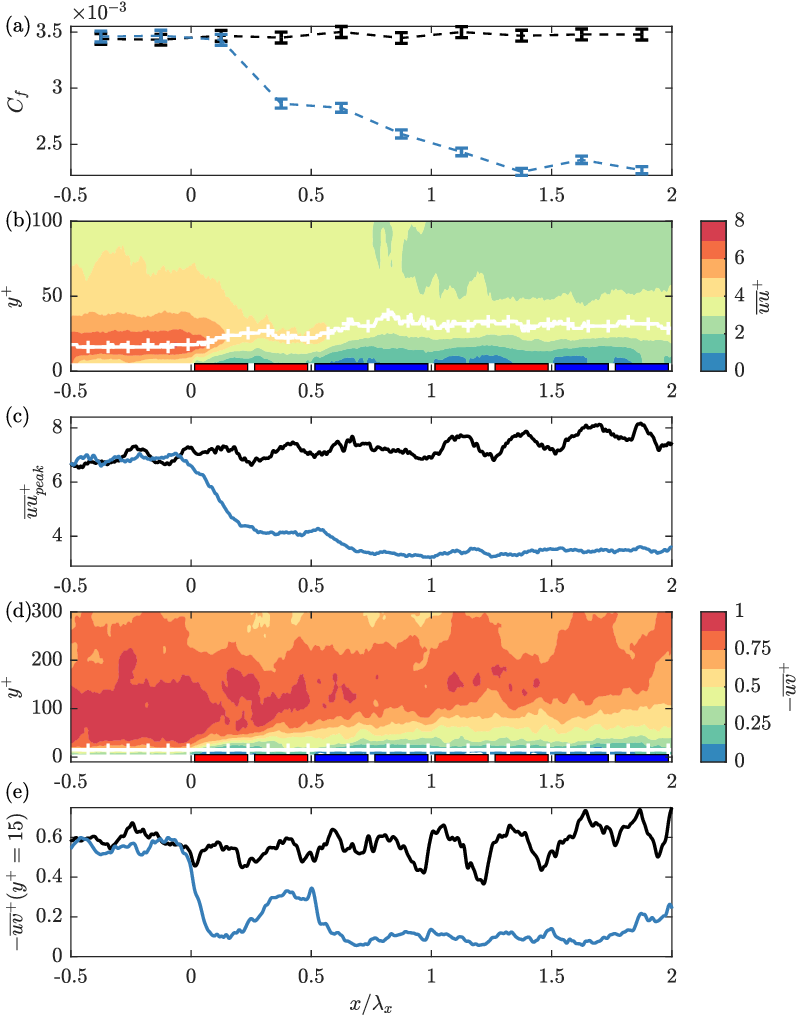}\\

    \caption{Streamwise evolution of (a) skin-friction coefficient ($C_f$), note only the markers correspond to data; dashed lines are added for interpretability, (b) $\overline{uu}^+$ and (d) $-\overline{uv}^+$ for spatial forcing of $\lambda_x^+ = 942$ and $A^+ = 12.0$, attained from FOV1. 
    The streamwise coordinate is scaled with $\lambda_x$ under the current conditions $\lambda_x \sim \delta_0$.
	White `+' markers in (b,d) respectively correspond to the line plots in (c,e), which compare $\overline{uu}^+$ and $-\overline{uv}^+$ between the actuated (in blue) and non-actuated case (in black).
    `+' marked in (c) corresponds to location of the near-wall peak in $\overline{uu}^+$, while that in (e) is fixed at $y^+$ = 15. Alike to figure \ref{fig:ExpSetup}, the shaded red (blue) regions denote the belts' location and positive (negative) spanwise motion direction, as such identifying the regions of forcing reversal.
		}
    \label{fig:FOV1_contours}
\end{figure}

We compare the skin-friction coefficient for the SqW and the non-actuated cases in figure~\ref{fig:FOV1_contours}(a) to quantify the rate of saturation of the drag modification. 
Here, we derive the nominal $C_f$ from the local values of $U_\tau$ used for the scaling in figure~\ref{fig:Umean_x}, based on the modified composite fit procedure (appendix~\ref{app:profileFit}). 
In line with the aforementioned behavior of the mean velocity profiles in figure~\ref{fig:Umean_x}, $C_f$ rapidly attenuates across the interval $x$ $\lesssim$ 1.5$\lambda_x$ $\sim$ 1.45$\delta_0$, after which it tends to saturate to a new reduced state. 
We note a nominal DR $\sim 30$\% at $x\sim 1.45\delta_0$, suggesting that the TBL undergoes a further slow modification as it approaches the downstream eight belts (FOV2; $9.7 \lesssim x/\delta_0 \lesssim 11.7$), where a nominal DR $\sim$ 38\% is obtained (see \S\ref{sec:DR}).

Investigation of the streamwise evolution of $C_f$ is supported by the analysis of the turbulence statistics in the overlying TBL, for which we consider contours of $\overline{uu}^+$ and $-\overline{uv}^+$ in the $x-y$ plane from FOV1 (figures~\ref{fig:FOV1_contours}b-e), scaling the streamwise coordinate with $\lambda_x$.
Note that, for this case, the wavelength is almost the same as the BL thickness; hence, scaling with these is interchangeable.
To enhance legibility without influencing the conclusions, a smoothening was applied to the statistical contours (also repeated later in figure~\ref{fig:FOV2_Cf_Rxx}). 
For this, a Gaussian kernel was chosen to preserve sharp features, with the images convolved using 9- and 15-pixel kernels for the two stresses, respectively.
Note that both these Reynolds stresses as well as wall-normal distances are normalized by $U_{\tau0}$, and hence this figure allows differences between actuated and non-actuated cases in the `absolute' sense to be appreciated \citep{Agostini_spanwise_2014,agostini_turbulence_2015,deshpande2024near}.
White lines with `+' markers indicate wall-normal location of the peak of $\overline{uu}^+$ in figure~\ref{fig:FOV1_contours}(b), while they indicate $y^+ = 15$ in figure~\ref{fig:FOV1_contours}(d), both of which are contours associated with the actuated case, ${\lambda}^+_x = 942$.
The magnitudes of $\overline{uu}^+$ and $-\overline{uv}^+$ at these locations are plotted in figures~\ref{fig:FOV1_contours}(c,e) respectively, for visualizing the spatial modifications of the near-optimal actuated case (in blue) against the non-actuated case (in black). 
As remarked previously in \S \ref{sec:setup}, the near-wall statistics display some spatial modulation, which is related to the actuation surface's imperfections; however, these do not essentially affect our conclusions.
It can be observed that the peak in $\overline{uu}^+$ is attenuated by $\sim$40\% and shifted to a higher wall-normal location almost immediately, already over the first belt (figures~\ref{fig:FOV1_contours}b,c).
The same can be noted for $-\overline{uv}^+$ in figures~\ref{fig:FOV1_contours}(d,e), which drops much more rapidly by $\sim 75-80\%$ within $\sim 0.1\delta_0$.
This strong attenuation signifies the dampening of the near-wall velocity streaks ($\overline{uu}^+$) that are inherently linked to the Reynolds shear-stress-producing events and, thereby, to the wall-shear stress \cite{jimenez1999autonomous,kim2011physics}. 
Along with the `quick' (\emph{i.e.}, on short length-scales) response, we can observe a more gradual attenuation in $-\overline{uv}^+$ that propagates deeper into the logarithmic layer (to beyond $y^+ > 300$), over a much larger streamwise extent covering at least the current FOV $x/\delta_0\gtrsim 2$.
This is a trend that qualitatively aligns with the established evolution of $C_f$.


Interestingly, however, as the TBL further evolves, both Reynolds stresses in the near-wall region start recovering to higher turbulence intensities as the TBL evolves towards the half-phase ($x$ $\sim$ 0.5$\lambda_x$).
While this recovery is relatively minor for $\overline{uu}^+$ and effectively `plateaus', it is significant and clearly noticeable for $-\overline{uv}^+$.
In contrast to $\overline{uu}^+$, the Reynolds shear stress is dominated by relatively small spatial and temporal scales, owing to which the latter responds more `quickly' to changes in the wall actuation condition \cite{sillero2014two}.
This plausibly explains the differences between the spatial responses of the two stress components evident in figure~\ref{fig:FOV1_contours}. 
For completeness, we make a note here that contours of $\overline{vv}^+$ are dominated by similar scales as to $-{\overline{uv}^+}$ motions, and we have avoided showing them here for brevity (all available Reynolds stress contours are provided in a supplementary dataset; see details in the data availability section).

\begin{figure}[t]
    \centering
    \includegraphics{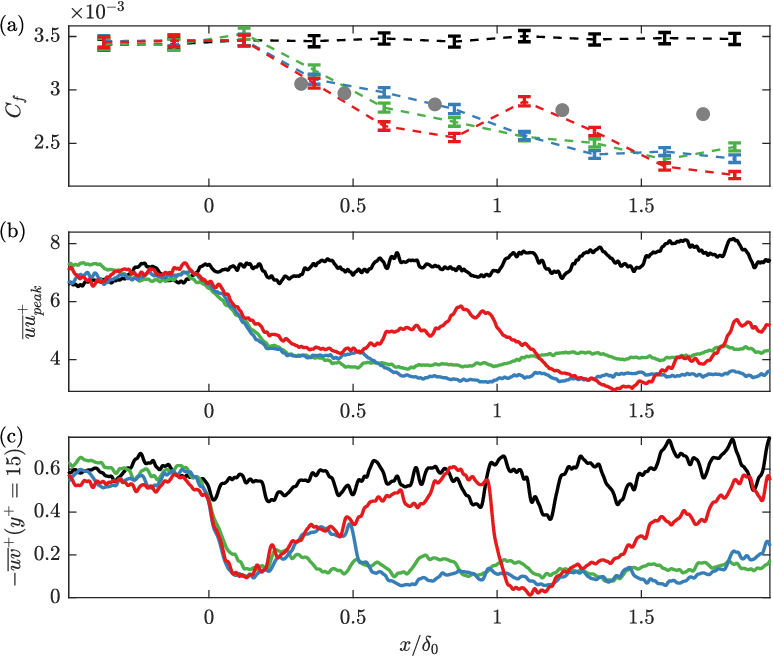}
    \caption{
    Streamwise evolution of (a) skin-friction coefficient ($C_f$), note only the markers correspond to data; dashed lines are added for interpretability, (b) $\overline{uu}^+$ and (c) $-\overline{uv}^+$ for wavelengths $\lambda_x^+ = $ 471 (\lineStyleAshort), 942 (\lineStyleBshort), 1884 (\lineStyleCshort) at $A^+ = 12.0$, and the non-actuated reference (\lineStyleRefshort), at the initiation of actuation (FOV1). In (a), the grey dotted markers indicate DR evolution of \citet{ricco_effects_2004}; skin friction was obtained following $C_f = C_{f0}(1-\mathrm{DR}/100)$, with $C_{f0}$ used of the current study.  
    The line plots correspond to the Reynolds stress at (b) the near-wall peak in $\overline{uu}^+$, while that in (c) is fixed at $y^+$ = 15. 
    }
    \label{fig:FOV1_linescomparison}
\end{figure}

We further investigate the recovery within the forcing phase (\emph{i.e.}, intra-phase recovery) by considering also the other two wavelength cases, $\lambda_x^+ = 471$ and 1884 (\emph{i.e.}, 0.5${\delta}_0$ and 2${\delta}_0$, respectively). 
Figure~\ref{fig:FOV1_linescomparison} presents the spatial evolution of $C_f$, and the line plots of the Reynolds stresses at similar $y^+$ locations as in figure~\ref{fig:FOV1_contours}, but now for all three cases. 
Note that in this figure, the $x$-coordinate is scaled with $\delta_0$, to better compare the development in physical distance, irrespective of the forcing wavelength. 
Figure~\ref{fig:FOV1_linescomparison}(b) confirms that the recovery of $\overline{uu}^+$, which was only weakly apparent in figure~\ref{fig:FOV1_contours}(c), is indeed physical and is much more clearly evident when imposing a larger $\lambda^+_x = 1884$ (in red). 
Both $\overline{uu}^+$ and $-\overline{uv}^+$ get `re-attenuated' at their half-phase location ($x\sim$ 0.5${\lambda}_{x}$) for all cases, where the actuation initially changes forcing direction (\emph{i.e.}, $x/\delta_0 \lesssim 0.25$, $x/\delta_0 \lesssim 0.5$, and $x/\delta_0 \lesssim 1$ for $\lambda^+_x = $ 471, 942 and 1884, respectively) 
Notably, we can observe a strong similarity in the manner in which stresses react to the initial forcing input, regardless of wavelength, before the reversal in the wall-forcing direction occurs.
Following its initial transient, the two cases for $\lambda_x^+ \lesssim 1000$, start to establish a  quasi-streamwise-homogeneous response, while the post-optimal
exhibits a strong phase-wise variation. 
Such a streamwise alternate variation between attenuation and recovery continues to be maintained further downstream, even when the forcing is fully established, which we detail further in \S \ref{sec:downstream}.

Assessing the skin-friction evolution in figure~\ref{fig:FOV1_linescomparison}(a), it is worth highlighting that even though the established DR for the cases $\lambda_x^+ = 471$ and 942 have a significant difference in magnitude (see \S \ref{sec:DR}), their initial spatial development ($x/\delta_0 \lesssim 0.5$) is very similar.
In fact, the $C_f$ trend for both these cases follows a similar response to the oscillating-wall experiment ($T^+ = 67$, $A^+ = 11.3$) of \citet{ricco_effects_2004} (in grey circles), hinting at universality in the initial TBL response for sub-optimal and near-optimal cases.  

We find that for the post-optimal case ($\lambda_x^+ = 1884$), and initially also the near-optimum case ($\lambda_x^+ = 942$), an interesting phenomenon occurs; namely, the response of $C_f$ and turbulence recovery are notably out-of-phase.
Over the first half-phase (\emph{i.e.}, $x<\lambda_x/2$), the skin friction reflects a monotonic decline while both $\overline{uu}^+$ and $-\overline{uv}^+$ display a significant recovery. 
Moreover, once the reversal of the forcing direction occurs for the post-optimal case, $C_f$ actually increases while $-\overline{uv}^+$ reveals a strong attenuation, similar to the one at initial forcing ($x = 0$).

Based on these findings, we can conclude that the majority of the flow transient for the (sub-)optimal cases occurs within $x/\delta_0 \lesssim 1$, after which there is only minor streamwise variation before the fully established control regime is attained (see further discussion in \S\ref{sec:downstream}). 
Hereby, we find that the present scaling of the streamwise transient with $\delta_0$ is consistent with the available literature, which shows that the transient behavior is insensitive to the actuation parameters as well as the Reynolds number (noted previously in \cite{Skote_wall_2019, skote_comparison_2013}).
Our findings for post-optimal wavelengths reflect a significant attenuation of turbulence activity, followed by its intra-phase recovery, in the initial and latter parts of the half-phase, respectively. 
Notably, we found that the intra-phase turbulence recovery and skin-friction evolution are strongly out-of-phase.
We aim to reconcile and discuss these findings more elaborately in \S\ref{sec:mechanisms} after we establish the associated phase-wise variations of the Stokes strain rate for SqW forcings in \S\ref{sec:analyticalSSR} (as previously established for SinW forcings in \S\ref{sec:intro_mech}).
 
\subsection{Distinguishing Stokes strain between sinusoidal and square-wave forcing}
\label{sec:analyticalSSR}

This subsection draws inspiration from the previous studies of Leschziner and co-workers \citep{Touber_near-wall_2012,Agostini_spanwise_2014,agostini_turbulence_2015} by determining the Stokes strain and the Stokes strain rate (SSR) for the SqW, in comparison to the SinW.
By qualitatively contrasting the characteristic SSR responses, we aim to further elucidate the observed different regimes of attenuation and recovery of turbulence over the half-phase of SqW wall-forcing (see \S\ref{sec:spatial_statistics}).
These studies were reviewed in \S\ref{sec:intro_mech} and describe the correlation between the occurrence of high SSR and near-zero SSR (\emph{i.e}, constant `lingering' Stokes strain) with turbulence attenuation and its recovery, respectively.
Although these past works were focused on temporal SinW forcings, where the SSR is defined as ${\partial^2 \widetilde{W}}/{\partial t \partial y}$, it is reasonable to conjecture that their findings also extend to spatial SinW forcing \citep{viotti_streamwise_2009,yakeno_spatio-temporally_2009}, for which equivalently the SSR would be defined by ${\partial^2 \widetilde{W}}/{\partial x \partial y}$. 

\begin{figure}[h!]
    \centering    \includegraphics{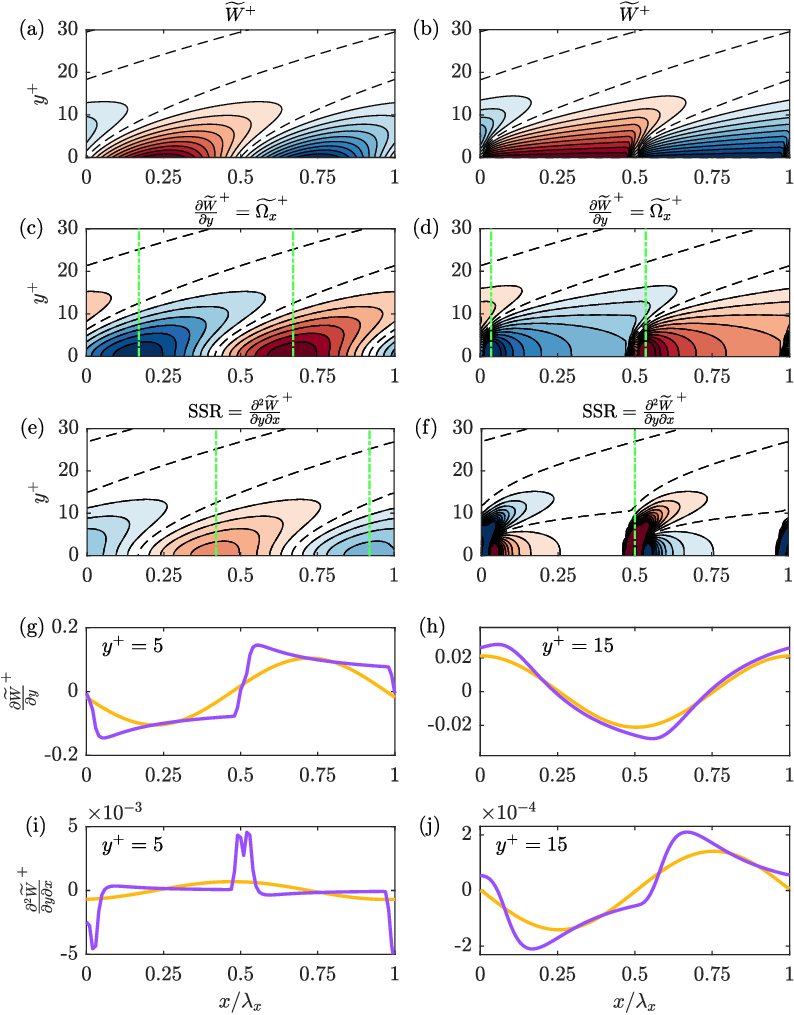}

    \caption{
    Phase-wise contours of analytical solution for the (modified) Stokes layer for (a,c,e) SinW \citep{stokes_effect_1851} and (b,d,f) SqW forcing \citep{knoop_experimental_2024}. 
    (a,b) Spanwise velocity, contour level range: -1:0.1:1; 
    (c,d) Stokes strain, ${\partial}\widetilde{W}^+/{\partial}y$ contour level range: -0.2:0.02:0.2; 
    (e,f) Stokes strain rate, ${{\partial}^2}\widetilde{W}^+/{{\partial}x}{{\partial}y}$ contour level range: -0.002:$\num{2e-4}$:0.002. Vertical green dash-dotted lines in (c-f), denote the locations of the respective wall-based peak magnitudes.
    We contrast the SinW (orange) and the SqW (purple) forcings by their respective (g,h) Stokes strain and (i,j) SSR, at (g,i) $y^+ = 5$ and (h,j) $y^+ = 15$.}
    \label{fig:SSL}
\end{figure} 

For the present discussion, we have chosen to deploy the modified spatial Stokes layer (SSL) model, considering unitary $A^+$ (note that the results are linear in $A$), which has been previously reported and validated using experimental data in \citet{knoop_experimental_2024} and also discussed sumarized in appendix~\ref{app:SSL} for completeness. 
The choice for using the validated SSL model was made in view of (i) the absence of stereoscopic PIV data required to compute the SSR directly, and (ii) the large uncertainties associated with the estimation of spatial gradients (especially second derivatives owing to measurement noise), even if stereo-PIV were conducted. 
The spatial wave characteristics for the modified SSL are selected to match the optimum DR case of $\lambda_x^+ = 942$. 
For this, we have verified that the 2-mm `static' regions between belts with the same $W_w$ sign do not significantly alter the global response \citep{knoop_experimental_2024}, and hence we have excluded this detail from the model for purposes of simplification, and treat adjacent belts running in the same direction as a single belt (further discussion to justify this approximation can be found in appendix~\ref{app:SSL}).
Figures~\ref{fig:SSL}(a,b) present the analytically obtained Stokes layer forcings ($\widetilde{W}$) for SinW (left) and SqW (right) scenarios across a single phase ($0 \leq x \leq \lambda_x$).
For reference, a three-dimensional representation of the SinW scenario was presented previously in figure~\ref{fig:SSL_strain}. 
The differences between SqW and SinW forcings can be clearly observed in the near-wall region, with the former having constant spanwise wall velocity magnitudes, except near 0.5${\lambda}_x$ and ${\lambda}_x$, which mark the location in-between the counter-running belts. 
However, the two strategies exhibit much more similar $\widetilde{W}^+$ behavior further away from the wall, likely due to the viscous diffusion effects. 

Next, we compare the Stokes strain (${\partial}\widetilde{W}/{\partial}y$) contours in figures~\ref{fig:SSL}(c,d) to the spanwise velocity contours. 
In the case of SinW forcing, the maximum Stokes strain region precedes the location of maximum spanwise velocities $-$ $x/\lambda_x=0.25$ and 0.75 $-$ by about one-eighth of the phase.
In contrast, the highest amplitudes of ${\partial}\widetilde{W}/{\partial}y$ for SqW forcing are obtained immediately downstream of the switch in the sign of $W_w$. 
We denote the location of the wall-based Stokes strain maxima by the dashed vertical lines, in the respective plots for SinW and SqW. 
The differences between ${\partial}\widetilde{W}/{\partial}y$ for SqW and SinW forcings are also compared quantitatively via a line plot at $y^+ = 5$ in figure~\ref{fig:SSL}(g). 
The peak magnitude for the SqW is nearly 1.5 times higher than for SinW (assuming the same forcing amplitude), and the aforementioned differences in phase-wise location of the peak strain are also evident. 
However, we observe a similar response between the two forcing types further away from the wall, which is confirmed by the line plot at $y^+ = 15$ figure~\ref{fig:SSL}(h).

This paves the way for investigating the SSR, which has been described as key to enforcing drag-reduction within a forcing phase (refer \S\ref{sec:intro_mech}).
By generalizing from the findings of Agostini and co-workers \citep{Touber_near-wall_2012, Agostini_spanwise_2014, agostini_turbulence_2015}, we pose the hypothesis that upon the imposition of high SSR, the turbulence is strongly attenuated, while an extended fetch of lingering Stokes strain can be associated with the recovery of turbulence.
As illustrated in figures~\ref{fig:SSL}(e,f) for SinW and SqW forcings, regions of elevated SSR magnitude correspond to the phases associated with a switch in the sign of $\widetilde{W}$, thereby occurring twice per period. 
However, the SSR is very small in the domain of maximum strain for SinW forcing, \emph{i.e.}, the strain lingers at a relatively low level in this region. 
The qualitative response for the SqW forcing, however, is markedly different.  
The forcing is observed to be much stronger, and is primarily `localized' to the region associated with a switch in the sign of $\widetilde{W}$ (\emph{i.e.}, forcing direction reversal). 
Hence, the dominant forcing term is strong and impulsive rather than the gradual variation noted in the case of SinW, which is reflected quantitatively in figures~\ref{fig:SSL}(i,j).
The peak SSR magnitude for SqW forcing is over 12 times stronger than that for SinW at $y^+$ = 5. 
This suggests that the local forcing experienced by the flow, which is responsible for attenuating the self-sustaining near-wall cycle, is one order stronger for the SqW than for the SinW.
Note, however, that this does not linearly correlate to DR, per its saturation under high amplitudes of forcing \cite{quadrio_critical_2004}.
Downstream of the high SSR forcing, the rest of the half-phase is characterized by lingering Stokes strain, \emph{i.e.}, near-zero SSR.
In \S\ref{sec:mechanisms}, we aim to associate these SqW-specific SSR characteristics directly with the response of turbulence (recovery), while we speculate plausible reasons for the out-of-phase behaviour noted for the skin-friction.
.

\subsection{Distinguising the response of turbulence and skin friction to square-wave forcing}
\label{sec:mechanisms}

\begin{figure}
    \centering
    \includegraphics{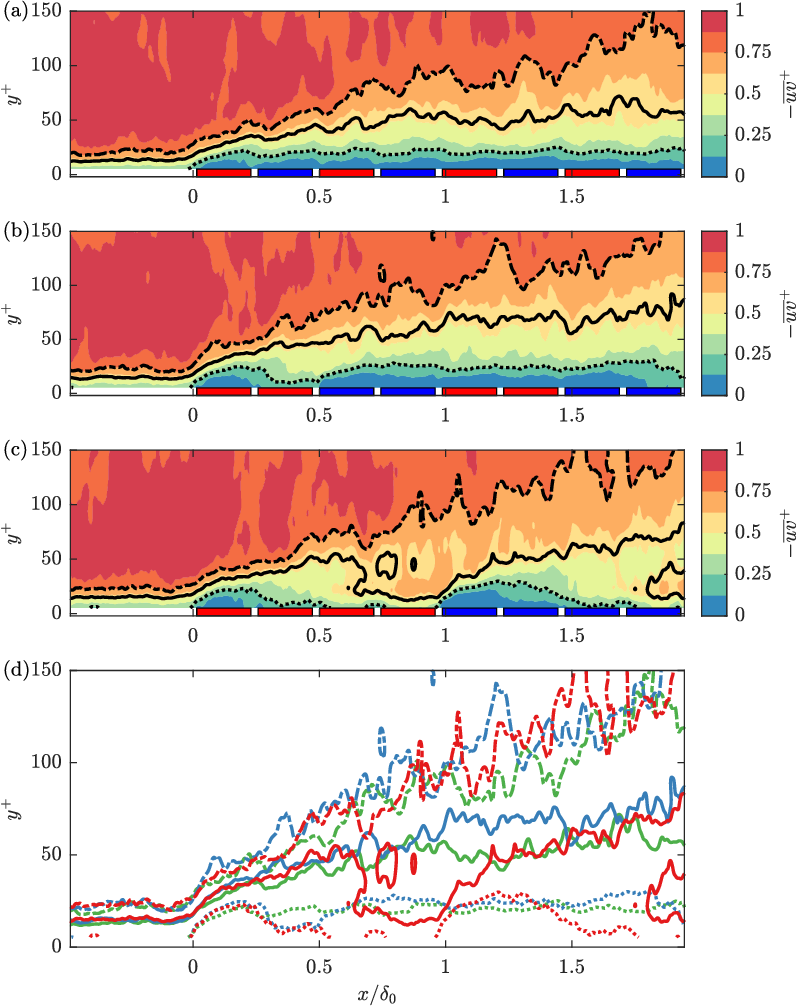}
    \caption{
    Streamwise evolution of $-\overline{uv}^+$ in the streamwise-wall-normal plane at the initiation of forcing (FOV1), for wavelengths (a) $\lambda_x^+ = $ 471 (\lineStyleAshort), (b) 942 (\lineStyleBshort), and (c) 1884 (\lineStyleCshort) at $A^+ = 12.0$. (d) presents a comparison of the black contour lines in (a-c) denote $-\overline{uv}^+ = [0.25\ 0.55\ 0.75]$ (dotted, solid, dashed). 
    Note the change in the limits of the wall-normal extent to $y^+ = [0,\ 150]$. Alike to figure \ref{fig:ExpSetup}, the shaded red (blue) regions denote the belts' location and positive (negative) spanwise motion direction, as such identifying the regions of forcing reversal.
    }
    \label{fig:FOV1_unContours}
\end{figure}

From our previous discussion, we can establish that the SqW has two clearly distinguishable regions of SSR within a half-phase ($\lambda_x/2$): sub-phase-I of high impulsive SSR associated with the reversal of forcing direction, followed by a sub-phase-II characterized by an extended domain of lingering Stokes strain (\emph{i.e.}, negligible SSR) owing to constant $W_w$. 
\revision{Here, we define sub-phases I and II based on the SSR at the wall to be significant and near-zero, respectively, which subsequently determines their streamwise fetch.}
Based on our analytical assessment of the near-optimum wavelength in figure~\ref{fig:SSL}, both sub-phases span approximately half of the half-phase (\emph{i.e.}, $\lambda_x/4$).
As a result of the SqW-type boundary condition, the extent and SSR magnitude of sub-phase-I is largely invariant with $\lambda_x$. In contrast, the fetch of lingering Stokes strain of sub-phase-II extends in proportion to the wavelength.
In this section, we aim to reconcile these two SSR sub-phases with the response of turbulence observed in \S \ref{sec:spatial_statistics}.
We also distinguish between the response of the turbulence and the friction drag to wall-forcing.

The starting point for our discussion is the streamwise-wall-normal contour of $-\overline{uv}^+$, at the initiation of forcing, where the intra-phase recovery is most pronounced. 
\revision{The choice for considering the Reynolds shear stress was motivated by its link to $C_f$ through a (weighted) wall-normal integration across the boundary layer \citep{fukagata2002contribution, elnahhas2022enhancement}.}
We report this through colored contours for all three $\lambda^+_x$ cases in figures~\ref{fig:FOV1_unContours}(a-c).
To facilitate the discussion on variation in intra-phase recovery with $\lambda_x^+$, we highlight three constant energy contour levels, $-\overline{uv}^+ = [0.25\ 0.55\ 0.75]$, which are subsequently overlaid in figure~\ref{fig:FOV1_unContours}(d) to compare and contrast their streamwise evolution. 
In the streamwise direction, an upward movement of the contours is representative of attenuation, while its movement toward the wall can be associated with turbulence recovery.
Upon initial imposition of forcing at $x = 0$, the aforementioned gradual streamwise evolution of an attenuated $-\overline{uv}^+$ state, penetrating high into the boundary layer to over $y^+ > 300$ (see figure~\ref{fig:FOV1_contours}d), is reflected among all cases.
This response suggests that the impulsive forcing generates an energy-attenuated internal boundary layer that subsequently propagates into the existing TBL, and through which it adapts to the new low-energy near-wall state initiated by the high SSR forcing.
Its streamwise evolution is best reflected by the collapse of the $-\overline{uv}^+ = 0.75$ contour for all cases throughout the considered streamwise extent.
Initially, for $x/\delta_0 \lesssim 0.5$, this similarity is also reflected in the $-\overline{uv}^+ = 0.55$ contour, after which the post-optimal case recovers, while the behavior of the two $\lambda_x^+\lesssim 1000$ cases stay similar throughout the streamwise extent.
As such, we propose, distilled from our experimental results, that an SSR forcing `event' during sub-phase-I carries with it a certain streamwise domain of influence, which extends significantly beyond the fetch of sub-phase-I, governed by a long-length scale of order $\delta_0$ (linked to the wall-normal propagation of the shear-stress attenuated state).
This trend is consistent with the $\delta_0$-scaling that we established for the evolution of $C_f$ upon initial forcing (figure~\ref{fig:FOV1_linescomparison}a). 

The significant intra-phase recovery observed in the case of $\lambda_x^+ = 1884$, best characterized by the $-\overline{uv}^+ = 0.25$ and $0.55$ contours, remains confined to the near-wall region ($y^+ \lesssim 75$) and can be associated with the lingering Stokes strain in the extended half-phase of constant $W_w$.
When the forcing direction reverses at $x/\delta_0 \sim 1$, and SSR is imposed again, a strong attenuation and a similar recovery region can be observed, emerging over the downstream region of the second half-phase.
This recovery phenomenon is suggested to be governed by a relatively smaller length scale as compared to the hypothesized $\delta_0$ for $C_f$, owing to the fact that it is established within the waveform.
With reference to the vortex tilting mechanism of \citet{agostini_turbulence_2015}, this phenomenon is likely to be linked to the recovery time of the streaks of approximately $t^+ \sim 50$ \cite{blesbois2013pattern}, which converts to a length scale of $\mathcal{L}^+ \approx \mathcal{U}_c^+ t^+  \sim 500$ considering the well-accepted near-wall convection velocity $\mathcal{U}_c^+$ $\sim$ 10.
This likely explains the observation of a significant intra-phase recovery notably for the post-optimal case, where the half-phase extends far beyond $\lambda_x^+/2 \gg 500$.

Based on this, we can hypothesize the wavelength-dependent response of the TBL as resulting from a balance between having a sufficient fetch for sub-phase-I forcing effect to establish its full potential, while limiting at the same time the intra-phase recovery associated with an extended sub-phase-II.
Attesting to this, in figure~\ref{fig:FOV1_linescomparison}(b) the sub-optimal forcing, with only a limited fetch between the respective sub-phase-I forcings, attains a less-attenuated state of the streamwise stress peak (representative of the turbulent streaks), compared to the near-optimum, 
whereas the post-optimal case reflects a significant intra-phase recovery of the turbulent streaks as sub-phase-II extends significantly beyond $\mathcal{L}^+\gg 500$.

Next, we associate the response of $C_f$ with the characteristic SSR sub-phases, particularly for the post-optimal case, which is characterized by a marked out-of-phase response between $C_f$ and the recovering turbulence. 
Note that the present PIV experimental data do not provide enough fidelity to investigate the DR mechanisms through an energy or enstrophy budget analysis considered by \citet{Agostini_spanwise_2014, agostini_turbulence_2015,umair2023vorticity}.
Hence, here we focus on comparing the response of $C_f$ relative to the SSR and attempt to reconcile present findings with the literature.
A detailed examination of \citet{agostini_turbulence_2015} reveals a similar out-of-phase pattern to our experiment,
with $C_f$ locally increasing when SSR is high (\emph{i.e.}, sub-phase-I) followed by its subsequent decline as the stokes strain lingers (\emph{i.e.}, sub-phase-II).
Inferring from our earlier experimental observations, the latter trend may plausibly be explained by the differences in the streamwise extent over which the two phenomena respond.
Specifically, we conjecture that the $\delta_0$-scaled behavior of decreasing $C_f$ leads to a gradual decrease over the half-phase in response to the preceding sub-phase-I forcing, \revision{owing to the wall-normal propagation of an attenuated $-\overline{uv}^+$ internal boundary layer}.
\revision{Concurrent is the emergence of turbulence recovery, which takes place predominantly in the near-wall region $y^+ \lesssim 50$, thereby contributing only minimally to the enhancement of $C_f$. 
For scrutiny, we have verified that the $y^+>50$ region contributes in the order of $\sim80-90$\% to the Reynolds shear stress integral of both the \citet{fukagata2002contribution} and \citet{elnahhas2022enhancement} identities (regardless of the $(1-y/\delta)$ weighting for the former).
However, the intra-phase recovery is believed to exert a more pronounced effect on the marked increase in $C_f$ during sub-phase-I, which will be addressed in the subsequent discussion.}
We aim to substantiate this phenomenon by referring back to the earlier introduced (refer to \S\ref{sec:intro_mech}) vortex-tilting/stretching mechanism by \citet{agostini_turbulence_2015}. 
As noted by \citet{umair2023vorticity}, the SSR-induced tilting of the near-wall turbulent streaks is associated with a localized increase in $C_f$. 
Therefore, a greater recovery of the streaks leads to a significantly higher local increase in $C_f$.
Our experimental data in figure~\ref{fig:FOV1_linescomparison}(b) supports this interpretation. 
The streamwise Reynolds stress peak, characteristic of these streaks, shows significant re-establishment near the end of the half-phase ($0.5 \lesssim x/\delta_0\lesssim 1$). 
At $x/\delta_0 \sim 1$, the subsequent application of high SSR in sub-phase-I induces tilting of these partially recovered streaks. 
Hence, this out-of-phase $C_f$ behavior becomes distinctly observable only under post-optimal conditions, where substantial streak recovery occurs during the lingering half-phase.
These characteristic phenomena become more clearly evident when the wall-forcing has fully established, which we will discuss next in \S\ref{sec:downstream}.

\subsection{Streamwise evolution of turbulence statistics under fully established forcing}
\label{sec:downstream}

\begin{figure}
    \centering
    \includegraphics{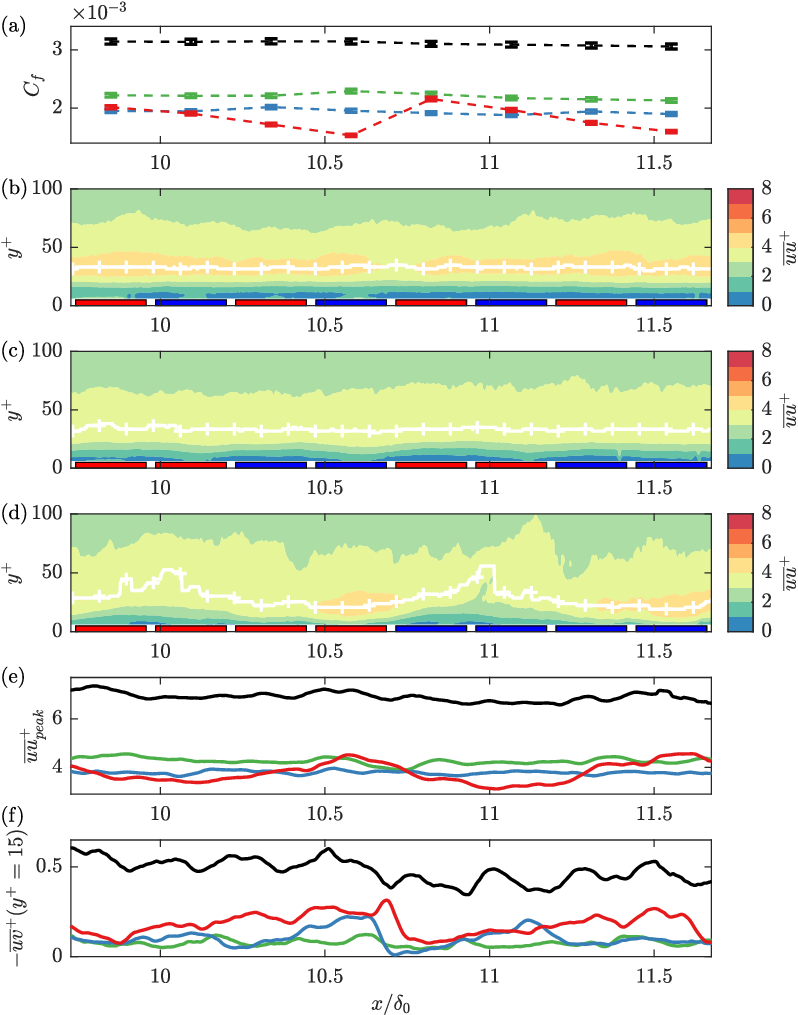}
    \caption{Streamwise evolution of (a) skin-friction coefficient ($C_f$), note only the markers correspond to data; dashed lines are added for interpretability, (b-d) $\overline{uu}^+$, for wavelengths $\lambda_x^+ = $ 471 (\lineStyleAshort), 942 (\lineStyleBshort), 1884 (\lineStyleCshort) at $A^+ = 12.0$, and the non-actuated reference (\lineStyleRefshort), towards the downstream end of forcing as obtained from FOV2.
    Similar to figure~\ref{fig:FOV1_contours}, the line plots in (e,f) correspond to the magnitude of the near-wall peak in $\overline{uu}^+$ (denoted by white `+' markers in (b-d)), and $-\overline{uv}^+$ at $y^+ = 15$, respectively. Alike to figure \ref{fig:ExpSetup}, the shaded red (blue) regions denote the belts' location and positive (negative) spanwise motion direction, as such identifying the regions of forcing reversal.
  }
    \label{fig:FOV2_Cf_Rxx}
\end{figure}

In this section, we examine the TBL response under fully established forcing conditions, further substantiating the previously discussed flow mechanics (\S\ref{sec:mechanisms}).
We are now concerned with 
the flow field over the most downstream region of the actuation domain (covering belts\# $40-48$), in FOV2.
As becomes evident from figure~\ref{fig:FOV2_Cf_Rxx}, which presents the streamwise variation of $C_f$, $\overline{uu}^+$, and $-\overline{uv}^+$, a fully developed regime is attained (\emph{i.e.}, no global spatial transient) for all three cases, while, at the same time, the post-optimal case still exhibits strong intra-phase variations.

Assessing the spatial variation of skin friction in figure~\ref{fig:FOV2_Cf_Rxx}(a) confirms streamwise homogeneous behavior for $\lambda_x^+ = 471$ and $942$, owing to the physically small streamwise spacing between the subsequent SSR forcing events (\emph{i.e.}, the half phase extends only $\lambda^+_x/2 \leq 500$), thereby not permitting significant recovery.
A similar effect is noted for both these cases in terms of the streamwise Reynolds stress (figures~\ref{fig:FOV2_Cf_Rxx}b,c), with the peak value location at $y^+ \approx 25-30$  relative to $y^+ \approx 15$ for the uncontrolled case (see also figure~\ref{fig:referenceBLprofiles}b). 
However, the turbulence attenuation is observed to be strongest for the optimum wavelength case. Accordingly, and in line with the available literature \cite{viotti_streamwise_2009,quadrio_streamwise-travelling_2009}, the near-optimum case ($\lambda_x^+ = 942$) outperforms the $\lambda_x^+ = 471$ case also in terms of skin-friction reduction (figure~\ref{fig:FOV2_Cf_Rxx}a). 
The established $C_f$ of both these actuated cases, relative to the non-actuated reference, reflects an integral DR $\sim32$ and $\sim38$\%, respectively (discussed elaborately in \S\ref{sec:DR}).

In accordance with the phenomenology discussed in \S\ref{sec:mechanisms}, the post-optimal case at  $\lambda^+_x = 1884$ (in red) reflects a qualitatively similar response; however, now fully established.
Upon the imposition of high SSR where forcing reversal occurs (most upstream and halfway through the FOV), the turbulent stresses are attenuated (figures~\ref{fig:FOV2_Cf_Rxx}d-f).
Subsequently, when the SSR lingers in sub-phase-II, turbulence recovery occurs, which is gradual for $\overline{uu}^+$ as compared to a relatively more impulsive response of $-\overline{uv}^+$.
Note that the magnitude of the intra-phase variation is not as high as in FOV1, likely owing to the TBL's overall energy-reduced state.
We also reaffirm the out-of-phase response of $C_f$ to the turbulence over the half-phase where $W_w$ is constant (sub-phase-II).
Here, $C_f$ is continually declining, whereas a local increase is observed upon reversal of forcing direction (\emph{i.e.}, high SSR, sub-phase-I).
This is supportive of our previously introduced inference that $C_f$ attenuates in response to the high SSR forcing over a longer spatial extent, while at the same time, the near-wall turbulence already recovers when the Stokes strain lingers.
Representative of the turbulent streaks, the streamwise stress peak in figures~\ref{fig:FOV2_Cf_Rxx}(d,e) exhibits significant phase-wise variations in magnitude, ranging between $\overline{uu}_{peak}^+ = 3.4-4.5$, and wall-normal location between $y^+ \sim 25-50$. 
The re-establishment of $\overline{uu}^+$ toward the end of the half-phase, at $x/\delta_0 \sim 10.5, \&\ 11.5$, is supportive of the SSR vortex-tilting mechanism to locally increase $C_f$ (discussed previously in \S\ref{sec:mechanisms}). 

To summarize, we have shown that the turbulence response in the fully-establsihed state is self-consistent with the previously discussed phenomenology in the initial transient region (\S\ref{sec:mechanisms}), which was associated with individual SSR sub-phases. 
A conceptual synthesis of the interpretation of these results will be presented and discussed in the section in \S\ref{sec:discussion} after we elaborate on the integral performance parameters of the SqW-type forcing in \S\ref{sec:DR}.

\subsection{Turbulent drag reduction via steady spatial square-waves of spanwise velocity}
\label{sec:DR}

\begin{figure}[t]
    \centering
    \includegraphics{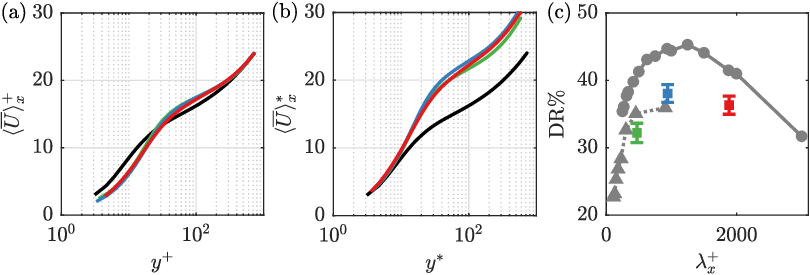}
    \caption{
    (a,b) Mean streamwise velocity profiles for wavelengths $\lambda_x^+ = $ 471 (\lineStyleAshort), 942 (\lineStyleBshort), 1884 (\lineStyleCshort) at $A^+ = 12.0$, and the non-actuated reference (\lineStyleRefshort), normalized by flow properties associated with (a) non-actuated ($U_{\tau0}$) and (b) actual ($U_{\tau}$) flow conditions.
(c) Drag reduction (DR\%) as a function of $\lambda^+_x$ for the SqW forcing scenarios investigated experimentally (in colored symbols), at $A^+ = 12$. 
Also considered for comparison are DR\% estimates from channel DNS considering SinW forcing: \lineStyleViotti \citet{viotti_streamwise_2009} at $Re_\tau = 200$ and \lineStyleGQ \citet{gatti_reynolds-number_2016} at $Re_\tau = 906$. 
}
\label{fig:DR}
\end{figure}

To conclude our discussion of the results, this subsection establishes the integral flow control potential of the SqW forcing imposed experimentally at $Re_{\tau} = 960$, for varying forcing wavelengths (table \ref{tab:experimental_overview}). 
A combination of ensemble as well as streamwise averaging is applied to the streamwise velocity profiles over the last eight belts in FOV2 (${{\langle}\overline{U}{\rangle}}_x$), and they are plotted after normalizing with the reference (non-actuated) friction velocity ($U_{\tau0}$) in figure~\ref{fig:DR}(a).
The profiles are averaged across four phases/forcing cycles (for $\lambda^+_x = 471$; in green), two phases ($\lambda^+_x = 942$; in blue) or a single phase ($\lambda^+_x = 1884$; in red) depending on the spatial forcing wavelength. 
However, these differences do not influence the variations exhibited by the ${{\langle}\overline{U}{\rangle}}_x$ profiles owing to the fully established control effect and minimal spatial development of the TBL in this region (previously verified in figure~\ref{fig:FOV2_Cf_Rxx}). 
The profiles, however, would be influenced by the streamwise extent of turbulence attenuation and recovery within each forcing phase, which varies with $\lambda^+_x$.

\begin{figure}[t]
    \centering
    \includegraphics{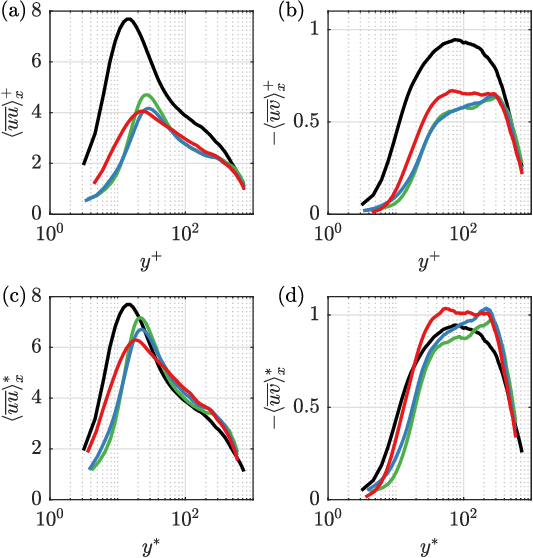}
    \caption{Wall-normal profiles of turbulence statistics: (a,c) streamwise Reynolds stress, and (b,d) Reynolds shear stress, scaled with (a,b) $U_{\tau0}$ and (c,d) $U_{\tau}$, respectively. 
		The linestyles correspond to the markers and the legend in fig.~\ref{fig:DR}, and the linestyle in table~\ref{tab:experimental_overview} for varying ${\lambda}^+_x$ cases.
  }
    \label{fig:BLprofiles}
\end{figure}

In line with the literature \citep{viotti_streamwise_2009,skote_comparison_2013}, all ${{\langle}\overline{U}{\rangle}}^+_x$ profiles for actuated cases collapse in the far outer region and exhibit a notable downward shift in the viscous sublayer.
The reduction of near-wall velocity is accompanied by an increase/overshoot of ${{\langle}\overline{U}{\rangle}}_x$ beyond $y^+ \geq 20$, which extends up to a wall-normal height of approximately $y^+ = 400$. 
This is a well-known feature \cite{Choi_mechanism_2001,quadrio_streamwise-travelling_2009,chandran_turbulent_2023} noted on imposition of spatio-temporal wall-forcing. 
\citet{Choi_mechanism_2001} explained that this modification results from a resultant mean spanwise vorticity component, $\widetilde{\Omega_z}$, due to the periodic spanwise tilting of the $\widetilde{\Omega_x}$ component associated with the Stokes layer.

We next implement the modified composite fitting procedure discussed in appendix~\ref{app:profileFit} on the mean velocity profiles, to estimate the nominal DR.
Figure~\ref{fig:DR}(b) depicts ${{\langle}\overline{U}{\rangle}}^*_x$ versus $y^*$ for the same cases as in figure~\ref{fig:DR}(a), but normalized by the actual $U_{\tau}$ estimated via the modified composite fit.
While the collapse of all ${{\langle}\overline{U}{\rangle}}^*_x$ profiles in the near-wall region is enforced, the vertical offset of the actuated cases in the intermediate/log-region (denoted by ${\Delta}B$) is a function of DR.
As expected, the ${{\langle}\overline{U}{\rangle}}^*_x$ profile associated with the $\lambda_x^+$ $\approx$ 1000 case has the highest vertical offset amongst the three wall-forcing cases, suggesting the greatest DR.
This is quantified in figure~\ref{fig:DR}(c) depicting DR as a function of $\lambda_x^+$ for all three SqW forcing cases considered at $Re_{\tau} = 960$. 
These estimates are compared against published results from the channel DNS of \citet{viotti_streamwise_2009} ($Re_\tau = 200$; purely spatial wall-forcing) and \citet{gatti_reynolds-number_2016} ($Re_\tau = 950$; STW forcing at $\omega = 0$), both considering SinW forcing at $A^+=  12$.
A maximum DR $\sim38.1$\%, with a 95\% confidence interval of $\pm 1.31$\%, is found for the present experiments at $\lambda_x^+ = 942$, which compares well with DR noted for the optimum wavelength case in the literature \citep{viotti_streamwise_2009,yakeno_spatio-temporally_2009,gatti_reynolds-number_2016}. 
In contrast, the sub- and post-optimal forcing cases have a DR of $\sim32.1 \pm 1.44$\% and $\sim36.3 \pm 1.35$\%, respectively. 
Further, the rate of variation of DR with ${\lambda}^+_x$ is consistent with the literature, with DR increasing rapidly for sub-optimal wavelengths and decreasing gradually in the post-optimal region.
The slightly higher DR noted for present experiments than DNS (for SinW), at comparable $Re_{\tau}$ \citep{gatti_reynolds-number_2016}, may be either owing to the overall greater efficacy of the SqW forcing than the SinW forcing as elucidate by \citet{cimarelli_prediction_2013} for temporal forcing, or an artifact of the modified composite fit methodology.
Nonetheless, the present focus is on obtaining the qualitative trend of DR\% with $\lambda^+_x$ for SqW forcing, which is consistent with the literature.

Wall-normal profiles of the streamwise and ensemble-averaged Reynolds stresses are presented, scaled with either $U_{\tau0}$ or $U_{\tau}$ in figures~\ref{fig:BLprofiles}(a,b) and \ref{fig:BLprofiles}(c,d), respectively. 
The near-wall peak in ${\langle}{\overline{uu}}{\rangle}_x$ plotted in figures~\Ref{fig:BLprofiles}(a,c) signifies the intensity of the near-wall velocity streaks and their self-sustaining cycle \cite{kline1967structure}, while ${\langle}{\overline{uv}}{\rangle}_x$ is considered in view of its analytical links with the wall-shear-stress \citep{elnahhas2022enhancement}.
In general, all the actuated cases reveal a strong control effect represented by the attenuation of the Reynolds stresses in an absolute sense, with the near-wall ${\langle}{\overline{uu}}{\rangle}^+_x$ peak reducing by up to 47\%.
When normalizing with actual $U_{\tau}$, we observe a slight increase in $\langle \overline{uu} \rangle_x^*$ and $-\langle \overline{uv} \rangle_x^*$ in the outer-layer, which is consistent with the experimental findings of \citet{chandran_turbulent_2023} for STW forcing.  
For the two cases corresponding to $\lambda_x^+ \leq 1000$, the upward shift in the peak location for both ${\langle}{uu}{\rangle}^+_x$ and ${\langle}{uu}{\rangle}^*_x$ suggests the near-wall viscous cycle moves further away from the wall, causing the reduction in wall-shear-stress \citep{gatti_reynolds-number_2016,rouhi2023turbulent}. 
Beyond the optimum wavelength, \emph{i.e.}, $\lambda_x^+ = 1884$, however, the near-wall ${\langle}{uu}{\rangle}_x$ peak gets broader and moves closer to the wall again. 
This is likely an artifact of streamwise averaging across both $-$ attenuated and recovering streamwise variance profiles $-$ noted across two distinct portions over the forcing phase (figure~\ref{fig:FOV2_Cf_Rxx}).  
The same reasoning can be extended to explain the unconventional mean ${{\langle}\overline{U}{\rangle}}_x$ profile in the log-region for $\lambda_x^+ = 1884$ case (figure~\ref{fig:DR}b). 
Similar modifications to the mean and turbulent flow properties have been observed previously by \citet{rouhi2023turbulent} on imposition of STW forcings beyond optimal frequencies, but this study presents a phenomenological explanation for the same based on a phase-wise variation of the SSR (\emph{i.e.}, DR mechanism).

\section{Conceptual sketch of the flow phenomenology and discussion}
\label{sec:discussion}

\begin{figure}[t]
    \centering
\includegraphics{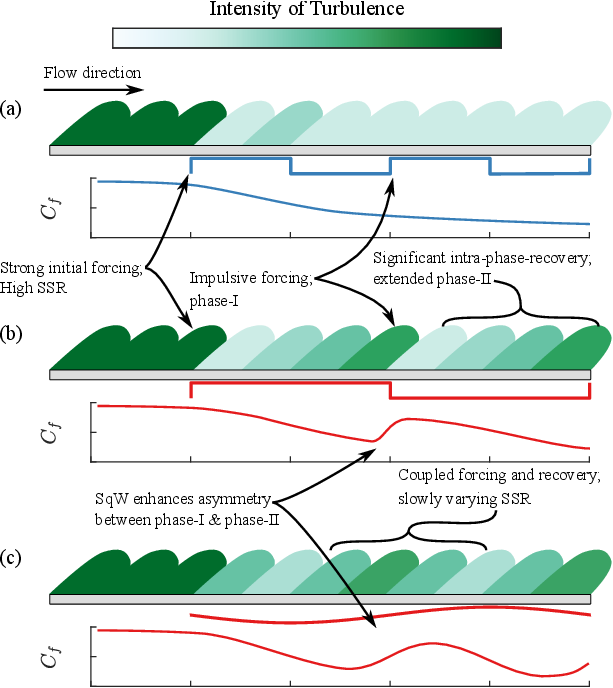}
    \caption{Conceptual sketch of the phenomenology illustrated schematically through the Reynolds shear stress turbulence intensity (in green shading) and $C_f$ (through line plots) for (a) optimal and (b) post-optimal SqW forcing, for which the two forcing phases and intra-phase recovery clearly establish. (c) shows post-optimum SinW forcing conceptualized based on past results of \citet{Agostini_spanwise_2014, agostini_turbulence_2015}, highlighting the more gradual and coupled phase-wise variation of turbulence attenuation and recovery compared to SqW forcing.}
    \label{fig:conceptualSketch}
\end{figure}

Based on the experimental results documented in the preceding sections, we have established the impact of wavelength variation on the streamwise evolution of the turbulence modification and  $C_f$ for SqW forcing. 
Founded on the available literature \citep{Bradshaw_measuements_1985,agostini_turbulence_2015}, we hypothesized that the SSR-related turbulence modification for SinW could be extended to the SqW, linking high SSR and lingering Stokes strain to respectively, turbulence attenuation and recovery.
For the SqW specifically, the SSR can be clearly distinguished into two regimes: (a) sub-phase-I of high and impulsive SSR that extends over a finite streamwise extent starting from the location of wall-forcing reversal (\emph{i.e.}, at $\lambda_x$/2) and quasi-independent of $\lambda_x$, and (b) sub-phase-II associated with near-zero SSR owing to constant wall velocity, leading to its region of influence being proportional to $\lambda_x$.
To clarify our observations to the reader, we present a conceptual sketch in figure~\ref{fig:conceptualSketch} of the TBL-modifying phenomenology, where we distinguish the modification in terms of wavelength and waveform. 
We present this sketch to represent the situation at the initiation of forcing since the response is most pronounced here, while at the same time, representative of the established dynamics that eventually emerge.
Our findings reflected the initial turbulence attenuation to be similar over the first half-phase of forcing, $x/\lambda_x<0.5$, for both the optimal and post-optimal cases as depicted in figures~\ref{fig:conceptualSketch}(a,b).
Such a response may plausibly be attributed to the similarity of the initially imposed SSR. 
Subsequently, the optimal case reflects a minor recovery over its half-phase until the successive imposition of SSR, after which a more-or-less established attenuated turbulence response emerges. 
It is important to highlight that a distinct response of the TBL to the individual SSR phases only emerges when the half-phase extends significantly beyond $\lambda_x^+/2 \gg 500$, \emph{i.e.}, post-optimal conditions.
For this case represented in figure~\ref{fig:conceptualSketch}(b), as sub-phase-II is extended, a significant intra-phase recovery of the near-wall turbulence becomes apparent.

For the post-optimal case, we observed the skin friction to reflect an out-of-phase response with respect to the turbulence attenuation.
In association with sub-phase-I, first a significant increase in $C_f$ could be observed, followed by its continual decline over the rest of the half-phase.
Although counter-intuitive at first, this type of behavior could plausibly be explained by the different length scales over which we found $C_f$ and turbulence to respond.
The intra-phase recovery was found to respond to a relatively short and viscous scale, which becomes particularly clear in the abrupt attenuation of $-\overline{uv}^+$ during sub-phase-I and its subsequent recovery.  
On the other hand, our experimental observations suggest skin friction responds to the forcing imposed low-energy state on a streamwise length scale of order $\delta_0$.
Moreover, the local increase in $C_f$ during sub-phase-I may be linked to the vortex-tilting mechanics \cite{agostini_turbulence_2015,umair2023vorticity} of the (partially) recovered turbulent streaks on the preceding half-phase. 
As a result of these dynamics, sub-phase-I induces a substantial attenuation of turbulence, accompanied by a significant jump in $C_f$; subsequently, over sub-phase-II, the skin friction is reduced in response to the imposed low energy state, while at the same time, the near-wall turbulence is already permitted to recover.

To discuss the impact of the forcing waveform, we compare the SqW and SinW at post-optimal forcing conditions in figures~\ref{fig:conceptualSketch}(b,c), where the latter is conceptualized based on the results of \citet{agostini_turbulence_2015}.
Qualitatively, our results on skin friction and turbulence attenuation corroborate their findings.
Interestingly, they noted an asymmetry between their drag-reduction phase and drag-increase phase, reminiscent of our sub-phase-II and sub-phase-I, respectively. 
As a consequence of the SinW forcing, however, these phases are not as distinct as the SqW but transition gradually from one to another, with both their regions of influence being proportional to $\lambda_x$.
Accordingly, for the SinW, the SSR magnitude is inversely proportional to $\lambda_x^+$, thereby effectively lowering the forcing strength with increasing wavelength/period.
Attesting to this, \citet{ding2023acceleration} have shown that the DR\% in temporal SinW forcing scales well with the acceleration, defined as $a^+ = A^+/T^+$.
On the other hand, the impulsive SSR response of SqW forcing enhances the asymmetry between the quasi-fixed sub-phase-I and $\lambda_x$-dependent sub-phase-II, so that the region of drag increase and turbulence attenuation is relatively short while the domain of $C_f$ reduction and recovering turbulence is extended.

\section{Concluding remarks}
\label{sec:conclusions}
We have presented an experimental study of wall-based transverse spanwise forcing by imposing a steady square-wave, with the objective of reducing the wall-shear stress in a turbulent boundary layer. 
A dedicated spatial forcing setup, with a streamwise extent covering $\sim 11.5 \delta_0$, has been implemented with the unique experimental ability to explore the TBL response to various independently controlled forcing parameters, \emph{viz.}, $\lambda_x$, $A$, and $Re_\tau$. 
Specific to our experimental implementation is the square-wave forcing discretization, achieved with a series of spanwise running belts. 
Hence, one of the objectives is to distinguish how the SqW impacts the forcing compared to the SinW waveform generally considered in numerical studies.
The present study focuses on the exploration of the streamwise evolution of a TBL at $Re_\tau = 960$ subjected to increasing forcing wavelength, namely $\lambda_x^+ = 471,\ 974,\ \mathrm{and}\ 1884$ at fixed $A^+ = 12$, in the streamwise fetch at the initiation as well as sufficiently downstream towards its fully established control state.

At the initiation of forcing, a strong control effect transient is established for all wavelengths considered, where a strongly reduced $C_f$ state is attained within $x/\delta_0 \leq 1.5$. 
Qualitatively, the streamwise variation in $C_f$ is consistent with the trends in the available literature for SinW forcing \cite{ricco_effects_2004, lardeau2013streamwise, skote_comparison_2013}, suggesting that $\delta_0$ is the relevant length scale given a universal collapse in the literature regardless of forcing type and actuation parameters.
Similar trends are also associated with the attenuation of $\overline{uu}^+$ and $-\overline{uv}^+$, most notable amongst which is the strong initial attenuation of $-\overline{uv}^+$ by $\sim 75 - 80\%$ over a streamwise extent of only $~0.1 \delta_0$.
When the wavelength increases beyond the generally recognized optimum of $\lambda_x^+ \approx 1000$ \cite{viotti_streamwise_2009}, to $\lambda_x^+ = 1884$, a significant recovery of turbulent stresses is observed within the forcing phase, with $-\overline{uv}^+$ re-establishing itself even towards its non-actuated stress levels.
We confirm that the intra-phase recovery is not dependent on the initial onset, as it is still significant for the post-optimum wavelength in the domain of fully established control ($\sim 9.5 - 11.5 \delta_0$). 
In this fully established domain, the flow modification is also characterized by the streamwise averaged turbulence statistics.
These agree with the available literature on (sub-)optimum forcing. 
The post-optimum case shows a broadening of the $\overline{uu}^+$ peak, which we attribute to averaging across significantly varying phase-averaged statistics, exposed to either turbulence attenuation or recovery.
Moreover, we established the control efficacy of the actuator with regard to skin-friction drag reduction by means of a modified composite fit of the streamwise-time-averaged velocity profile. 
Qualitatively, our DR variation with $\lambda_x^+$ aligns with the previous findings of \citet{viotti_streamwise_2009} and \citet{gatti_reynolds-number_2016}, with a maximum of $\sim38\%$ for the near-optimum forcing case.

The particular turbulence modification behavior of the SqW strategy is associated with its Stokes strain rate (SSR) topology, which allows for the identification of two markedly different phases; sub-phase-I associated with an impulsive imposition of high SSR, and sub-phase-II of prolonged lingering Stokes strain (\emph{i.e.}, near-zero SSR).
Based on the available literature, we hypothesized these regions to be linked to attenuation and recovery of the near-wall turbulence, respectively \citep{Agostini_spanwise_2014,agostini_turbulence_2015}.

To better understand the intra-phase recovery for SqW forcings relative to SinW, we explored the phase-wise variation of the SSR through a modified Stokes layer model \cite{knoop_experimental_2024}.
In the case of SinW forcing, the SSR varies slowly but continuously throughout the phase, reaching a maximum value where the wall velocity changes sign. 
In contrast, the SSR is imposed impulsively in the case of the SqW forcing (sub-phase-I), with its instantaneous magnitude found to be about one order higher than for the SinW forcing at $\lambda_x^+ \approx 1000$, for the same forcing amplitude.
This impulsively imposed SSR is concentrated at the $x$-location where the switch in the spanwise wall-forcing direction occurs, followed by an extent of negligibly small SSR (sub-phase-II) across the rest of the half-phase (owing to a constant spanwise wall velocity). 
For the SqW forcing, sub-phase-I, where the forcing direction reverses, is largely independent of wavelength, while sub-phase-II over the lingering half-phase extends in proportion to $\lambda_x$.
In essence, the SqW enhances the asymmetry, proportional to $\lambda_x$, between the observed drag-reduction and drag-increasing sub-phases established for SinW forcing by \citet{agostini_turbulence_2015}. 
This link between the lingering sub-phase-II with the intra-phase recovery of turbulence only establishes itself for post-optimal conditions when the half-phase extends significantly beyond $\lambda_x^+/2\gg 500$.
The out-of-phase response between $C_f$ and the near-wall turbulence may plausibly be explained by the different length scales on which the two phenomena respond, being $\delta_0$ and a viscous length, respectively.
Over the lingering sub-phase-II, while the turbulence is already recovering, $C_f$ declines/attenuates in response to the turbulence attenuation imposed by sub-phase-I.

The experimental work and established SSR-related phenomenology provide valuable foundations for understanding flow-modifying mechanisms. 
These insights can facilitate future studies and inform new hypotheses, particularly regarding the post-optimal regime, wherefore several key open questions arise. 
Firstly, as the SinW reduces in efficacy in proportion to $\lambda_x$ \citep{ding2023acceleration}, the question becomes as to how the intra-phase recovery and $C_f$ evolve over the lingering sub-phase-II when extending to an even higher wavelength regime.
An inquiry into the SqW and SinW performance at such conditions may become relevant in light of the outer-scaled actuation (OSA) strategy proposed and tested by Marusic and co-workers \citep{marusic_energy-efficient_2021,chandran_turbulent_2023,deshpande2024near}.
A downside of the SqW-type forcing is the significant theoretical power requirements associated with the extended domain of a constant spanwise wall velocity. 
As such, the efficacy of the control strategy could be enhanced through waveform optimization by setting part of the wall velocity to $W_w = 0 $ during the lingering sub-phase-II. 
Therefore, investigating the TBL's response to the (partial) absence of forcing, as opposed to the lingering Stokes strain, is considered valuable.
Moreover, the study's emphasis on SSR as a key driver of the drag-reduction mechanism may also offer valuable insights and support the development of future passive forcing implementations.

\begin{acknowledgments}
We would like to extend our special gratitude to the teams at Dimple Aerospace B.V. and BerkelaarMRT B.V. for their invaluable assistance in the development of the experimental apparatus. 
Additionally, we acknowledge and thank the technical staff of the wind tunnel laboratories at Delft University of Technology for their support.
The financial support of The Netherlands Enterprise Agency, under grant number TSH21002, and the funding acquisition by Dimple Aerospace B.V., 
are gratefully acknowledged by the authors. 
R. Deshpande gratefully acknowledges financial support from the University of Melbourne's Postdoctoral Fellowship.

\end{acknowledgments}

\section*{Author contributions}
M. Knoop: Conceptualization, Methodology, Investigation, Formal Analysis, Writing - Original Draft.
R. Deshpande: Conceptualization, Investigation, Writing - Review \& Editing, Supervision.
F. Schrijer \& B. van Oudheusden: Writing - Review \& Editing, Supervision.

\section*{Data availability}
\label{sec:Data}
A full dataset accompanying the individual figure will be made available \href{https://doi.org/10.4121/adb0b003-4912-4bdb-b12f-e56d18c3b937}{here} \cite{Dataset}, it contains the dimensional figure data as well as the relevant scaling parameter and atmospheric conditions. For completeness, we also provide the $x-y$ contours of $\overline{U}$ and $\overline{vv}$ accompanying figures~\ref{fig:FOV1_contours} and \ref{fig:FOV2_Cf_Rxx}.  

\appendix
\section{Wall shear stress determination through the modified composite fit}
\label{app:profileFit}
The well-established methodology of using a composite profile fit to the mean streamwise velocity profile, proposed by \citet{chauhan_criteria_2009}, is used here to obtain the mean friction velocity and the TBL thickness.
Their original formulation is a combination of the inner-profile description by \citet{Musker_explicit_1979} and an exponential formulation of the wake-profile based on high $Re_\tau$ measurements \cite{nagib_high_2006}. 
The composite profile considers log-law constants, $\kappa = 0.384$, $B = 4.17$ and is a function of the following variables: $U^+_{composite}(U_\tau, \nu, \Pi, \delta)$, where $\Pi$ is the wake parameter. 
Experimental studies are often subjected to an inherent uncertainty in the absolute wall-normal position of the first point from the wall \cite{orlu_near_2010}. 
This can be overcome by including a wall-normal offset ($\Delta y$) in the fitting parameters for the experimental data, as implemented previously by \citet{Rodriguez_robust_2015}.
They showed that the method performs accurately when the first datapoint is located $y^+\leq 10$, to a 95\% confidence interval for $U_\tau$ of $\pm 0.7\%$.
Furthermore, they detail the method's robustness to near-wall distortion of the first point from, for e.g., hot-wire conduction effects or biased near-wall PIV measurements. 
For the profiles that are considered in the work, spurious points, which were found up to a maximum of $y^+<4$, were manually removed before the final fitting procedure was performed.

\begin{figure}[t]
    \centering
    \includegraphics{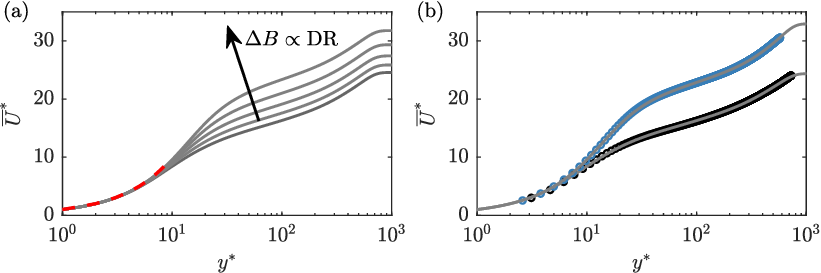}
   
    \caption{(a) Variation of the modified composite profile with DR\% ranging across 0:10:40. 
    Red dashed lines denote $U^* = y^*$. 
    (b) An example of the modified composite fit (in grey shading) compared against the experimental data discussed in \S~\ref{sec:DR} for the non-actuated (black) and actuated (blue) cases at $\lambda_x^+ = 942$, $A^+ = 12$.} 

    \label{fig:CompositeProf}
\end{figure}

The composite profile description is limited to canonical turbulent boundary layers. 
However, the formulations fall short if applied directly (\emph{i.e.}, without any modifications) to non-canonical cases, such as drag-reduced flows. 
To bridge this gap and quantify the DR\%, we propose modifying the existing formulation by incorporating the additive constant (${\Delta}B$) to the log-law, during the data fitting procedure. 
This is based on the observation \cite{gatti_reynolds-number_2016} that the mean streamwise velocity over an actuated surface, when scaled with the corresponding friction velocity, continues to exhibit inner-scaling in the viscous sublayer while experiencing an upward shift (${\Delta}B$) in the logarithmic layer.
We illustrate this in figure~\ref{fig:CompositeProf}(a) by showing the modified composite profiles (compared to the canonical formulation) for DR\% varying across 10:10:40, where $\overline{U}^*$ and $y^*$ denote normalization by the corresponding friction velocity. 
Visually, one can observe the modification as an extension of the buffer region and the characteristic upward shift in $\overline{U}^*$. 
We show the quality of the fit to our experimental data in figure~\ref{fig:CompositeProf}(b), depicting the reference case and actuated case from the discussion in \S~\ref{sec:DR}. 
The fitting procedure is implemented in an iterative fashion using a sequential quadratic programming method. The objective is to minimize the mean quadratic error (eq.~\ref{eq:error}) between the composite and experimental profiles. The experimental profiles are sampled on a logarithmic scale, where the vector pitch between the first points is retained, to give equal weighting to the inner- and outer layers.
\begin{equation}
    \label{eq:error}
    E = \overline{\sqrt{\left(\overline{U}^+_{composite}(U_\tau, \Delta y, B, \nu, \Pi, \delta) - \overline{U}^+(U_\tau)\right)^2}}.
\end{equation}
The uncertainty in $U_\tau$ for the modified fit was assessed using FOV 2 data for the $\lambda_x^+ = 472$ case (figure~\ref{fig:FOV2_Cf_Rxx}). With $S = 2$, each datapoint represents the same half-phase location, expecting consistent $U_\tau$ values. After detrending to remove minor streamwise variations, Students' T-distribution (i.e. eight samples) yielded a 95\% confidence interval of $U_\tau \pm 0.75\%$, aligning with the classical formulation's uncertainty.
. 

\section{The laminar solution for modified spatial Stokes layer}
\label{app:SSL}

The analytical solution of the spatial Stokes layer (SSL) for SinW forcing, a derivative from the second Stokes problem (\emph{i.e.}, temporal transverse oscillations) \cite{stokes_effect_1851}, was initially presented by \citet{viotti_streamwise_2009}:

\begin{equation}
\label{eq:SSL}
\widetilde{W}(x,y) = A C_x \Re \left[ e^{\mathrm{i} k_x x} \mathrm{Ai} \left(-\frac{\mathrm{i} y}{\delta_x} e^{-\mathrm{i}4\pi/3}\right)   \right],
\end{equation}
with $\delta_x = \left( \nu/(k_x u_{y,0})\right)^{1/3}$ as the penetration depth of the Stokes layer. 
In this relation, $\Re$ represents the real-valued component, $\mathrm{i}$ denotes the imaginary unit, $\mathrm{Ai}$ is the Airy function of the first kind, $C_x = \mathrm{Ai}(0)^{-1}$ is a normalization constant, and $u_{y,0}$ is the slope of the streamwise velocity profile at the wall. 
The Stokes layer solution has been validated extensively against both experimental and numerical data \cite{viotti_streamwise_2009, choi_near-wall_2002, deshpande2024near}. 
Given that we are imposing a different wall-boundary condition, specifically the square wave (SqW), the original solution for the SSL is no longer applicable. 
To address this, \citet{knoop_experimental_2024} introduced a modified SSL that uses a Fourier series to prescribe the desired periodic boundary condition. 
Due to the linearity of the governing $z$-momentum equation, the elementary SSL solutions can be summed. 
This approach is inspired by \citet{cimarelli_prediction_2013}, who applied a similar technique for temporal forcing. 
Following this formulation, the wall velocity is given as:
\begin{equation}
    W_w(x) = A\sum_{n=-\infty}^{+\infty} B_n e^{\mathrm{i} k_xn x}, 
\end{equation} where $n$ denotes the n\textsuperscript{th} Fourier mode with complex coefficient $B_n$. 
We now superpose these solutions to obtain the modified SSL model for an arbitrary waveform:
\begin{equation}
\label{eq:SSL_general}
\widetilde{W}(x,y) =
 A C_x \sum_{n=-\infty}^{+\infty} \Re \left[ B_n  e^{ik_x n x} \mathrm{Ai} \left(-\frac{i y}{\delta_x n^{-1/3}} e^{-i4\pi/3}\right)\right].
\end{equation}

Consistent with classical formulations, \citet{knoop_experimental_2024} demonstrated a good match between the modified SSL and experimental data. 
In its practical implementation, the SqW is convolved with a Gaussian function to prevent Gibbs phenomena, following a methodology similar to that used by \citet{gallorini2024spatial} for discrete STWs. 
A filter width of 1 mm was chosen carefully to minimize dispersive error while maintaining adequate spatial resolution. 

\begin{figure}[t]
    \centering
    \includegraphics{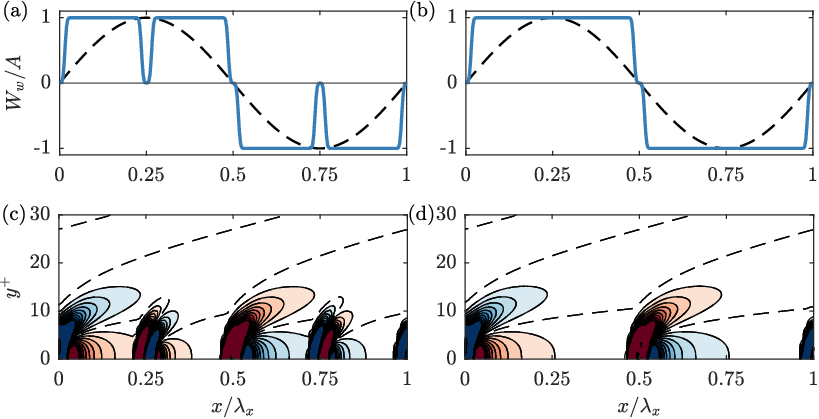}
    \caption{A comparison of the modified SSL between (a,c) the actual imposition of $W_w$, including 2-mm static, and (b,d) the simplified waveform. (a,b) imposed spanwise wall velocity for the case $\lambda_x^+ = 943$, with $S=4$ \#~belts, and (c,d) their respective SSR response. }
    \label{fig:SSL_disc_comparison}
\end{figure}

For our discussion in \S~\ref{sec:analyticalSSR}, the waveform for the cases with $S>2$ was simplified by removing the static 2-mm transition regions between belts that move with the same sign of $W_w$.
Figure~\ref{fig:SSL_disc_comparison} compares the two scenarios, where (a,b) highlights differences in wall-based spanwise velocity, the inclusion of the static transition can be appreciated on the left (a,c).
We compare the SSR, the primary quantity associated with the control effect in figures~\ref{fig:SSL_disc_comparison}(c,d), revealing an additional imposition of a respective positive and negative (and vice versa) SSR in (c).
Our results (\emph{e.g.}, figure~\ref{fig:FOV1_contours}) do not evidence a substantial impact of these static regions on the response of the TBL, with modifications only observed corresponding to a change in the $W_w$ sign. 
It is with this justification that we chose to simplify the wall velocity to better illustrate the discussion on the SSR.

\bibliography{bibliography}

\end{document}